\documentclass[11pt]{article}
\pdfoutput=1
\usepackage{jcappub,natbib}
\input{colordvi.tex}

\def\la{\mathrel{\raise.3ex\hbox{$<$\kern-.75em\lower1ex\hbox{$\sim$}}}}

\title{ET sensitivity to the anisotropic Stochastic Gravitational Wave Background}

\author[a]{Giorgio Mentasti,}
\author[a,b]{Marco Peloso}

\affiliation[a]{Dipartimento di Fisica e Astronomia “Galileo Galilei” Universit\`a di Padova, 35131 Padova, Italy}
\affiliation[b]{INFN, Sezione di Padova, 35131 Padova, Italy}

\abstract{We study the sensitivity of a pair of Einstein Telescopes (ET) (hypothetically located at the two sites currently under consideration for ET) to the anisotropies of the Stochastic Gravitational Wave Background (SGWB). We focus on the $\ell =0,2,4$ multipoles of an expansion of the SGWB in spherical harmonics, since the sensitivity to other multipoles is suppressed due to the fact that this pair of detector operates in a regime for which the product between the observed frequency and the distance between the two sites is much smaller than one. In this regime, the interferometer overlap functions for the anisotropic signal acquire very simple analytic expressions. These expressions can also be applied to any other pairs of interferometers (each one of arbitrary opening angle between its two arms) operating in this regime. Once the measurements at the vertices of the two sites are optimally combined, the sensitivity to the multipoles of the SGWB depends only on the latitude of the two sites, on the difference of their longitude, but not on the orientation of their arms. 
}

\begin{document}

\maketitle
\flushbottom

\section{Introduction}
\label{sec:intro} 

The Einstein Telescope (ET) is a proposed European ground-based gravitational-wave detector of third-generation which could be operating in the mid 2030s \cite{Punturo:2010zz,ET-Punturo}. Third generation (3G) detectors will have an order of magnitude improved sensitivity with respect to the current second-generation (2G) detectors (such as Advanced LIGO, Advanced Virgo, and KAGRA) and will span a greater frequency range. For ET, the improved sensitivity is due to a number of factors: the greater length of the arms ($10$ km, compared to $3$ km for Virgo and $4$ km for LIGO) will reduce displacement noises; placing the detector a few hundred meters underground will reduce gravity gradient and seismic noise; the presence of three nested detectors, placed at the vertices of an equilateral triangle will allow to resolve both GW polarizations, will provide a better antenna pattern, and will allow for better noise characterization through redundancies; finally, each detector will actually be composed by two interferometers, in what is denoted as a xylophone configuration, to better sample lower and higher frequencies. 

This improvement will allow us to address several key questions in astrophysics, cosmology, and fundamental physics \cite{Maggiore:2019uih}. For example, collisions between compact binaries with total mass in the range of $20-100$ solar masses will be visible by ET up to redshift $20$ and higher (as compared to the $z \simeq 1$ target sensitivity of 2G detectors), providing us with unprecedented information on the star-formation history, and probing the Universe before the birth of the fist stars (black hole mergers at such distances would necessarily have a primordial origin). As another example, the improved sensitivity and wider frequency coverage will allow for a better discrimination between different models of neutron stars composition, probing details of the merger and post-merger phenomena that are inaccessible to 2G detectors. Concerning cosmology, the high detection rate expected at ET, and its 
access to higher redshift than current detectors will allow for a  sub-percent level accuracy on the determination of the Hubble constant $H_0$, as well as more stringent tests on the dark energy equation of state, and, more in general, on modifications of General Relativity. 

Once correlated with other third generation detectors, as for instance Cosmic Explorer, the ET will also improve our sensitivity to the Stochastic Gravitational Wave Background (SGWB). The SGWB can have a cosmological and an astrophysical component. Among the cosmological sources, the amplification of quantum vacuum fluctuations during inflation is expected to be at an undetectable level for  3G detectors. However, several other mechanisms related to inflation could produce a detectable signal. 
\cite{Guzzetti:2016mkm,Bartolo:2016ami}. Other cosmological sources of the SGWB include pre-big-bang models, phase transitions, and topological defects (see \cite{Caprini:2018mtu} for a review). The astrophysical component originates instead from the superposition of a large number of unresolved sources that are too weak to be detected individually. In the frequency range probed by ground-based detectors, the strongest astrophysical SGWB is expected to be one due to the coalescence of black holes and neutron star binaries.

The most immediate step to disentangle the cosmological and astrophysical components of the SGWB is through the spectral dependence of its average (monopole) amplitude. Beside this, crucial information will be contained in its directionality dependence. The angular anisotropies (namely, the difference between the SGWB from any given direction, and the average monopole value) provide information about the angular distribution of the astrophysical sources \cite{SGWB-astro-angular} and might also become a tool to trace astrophysical or cosmological structures \cite{SGWB-astro-sources}. Anisotropies in the astrophysical background correlate with the Large Scale Structure distribution, due to both how the GW originate and on how they propagate to arrive to Earth \cite{SGWB-astro-correl}. Anisotropies in the cosmological component can also be inherent in their production mechanism 
\cite{Geller:2018mwu,Bartolo:2019zvb} or originate from the GW propagation in the perturbed cosmological background \cite{SGWB-cosmo,Bartolo:2019oiq}. This might also imprint a non-Gaussian statistics to the cosmological SGWB angular anisotropies, so that the SGWB might also be a new probe of primordial non-Gaussianity \cite{Bartolo:2019oiq}. To characterize the anisotropies, one typically decomposes the SGWB in  spherical harmonics $Y_{\ell m}$ (in one given chosen fixed cosmological frame), and then studies the correlation between different multipoles of this decomposition. 

Any measurement ($m$) of the SGWB is affected by noise ($n$) and, if present, by a signal ($s$). Denoting by ``1'' and ``2'' the measurements from two different detectors, the expected signal to noise ratio (SNR) is schematically of the form 
\begin{equation}
{\rm SNR} = \frac{\left\langle s_1  s_2 \right\rangle 
}{\sqrt{\left\langle \left( n_1 \, n_2 \right)^2 \right\rangle}}  = 
\frac{\left\langle m_1  m_2 \right\rangle - \left\langle n_1  n_2 \right\rangle 
}{\sqrt{\left\langle \left( n_1 \, n_2 \right)^2 \right\rangle}} \;.  
\end{equation} 
In this expression $\langle \dots \rangle$ denotes a correlation over time (to be defined in details below), 
and signal and noise are assumed to be uncorrelated from each other. At the numerator the expectation value of the noise has been subtracted from the measurement, to achieve an unbiased determination of of the signal, and in the denominator we have assumed that variance of the noise dominates that of the signal (weak signal regime). This ratio grows with the square root of the observation time, so assuming that the noise of the instrument is perfectly known (so that it can be perfectly subtracted at the numerator) one can in principle measure a signal much smaller than the noise if the observation time is sufficiently long. In practice, there will be a limit to how much the noise can be characterized. In particular, we cannot measure it in absence of a stochastic signal. Therefore, we cannot determine a signal below the level to which the noise is known. 

ET has individual detectors at the different vertices of its equilateral configuration. Measurement from these detectors are taken independently, but are not noise-uncorrelated, due of the proximity of the instruments. To cope with this problem, different and well apart ground-based interferometers are typically cross-correlated in searches for a SGWB. There are two sites currently under consideration for ET: one in the Sos Enattos mine in Sardinia, and one at the Belgium–Netherlands border. In this work we consider the sensitivity to the anisotropic SGWB of a hypothetical pairs of detectors placed at the two sites. We assume that the large separation, of about $1,200$ km, guarantees the the noise in one site is uncorrelated from that in the other site. Although we are not aware of a plan to construct two copes of ET at the two sites, we make this choice for simplicity, assuming that the two instruments are identical to each other, and that they have  the same specifications that are under consideration for ET. 

The observation of the anisotropic SGWB has been the object of several studies concerning GW earth-based \cite{Allen:1996gp,anisotropy-ground}, space-based \cite{anisotropy-space} interferometry and PTA \cite{anisotropy-PTA}. In particular, ref. \cite{Allen:1996gp} developed a formalism for the response to the anisotropic SGWB of detectors that are bound to the surface of the Earth, and that therefore have a regular scan pattern related to the daily rotation of our planet. At the core of this formalism is the computation of overlap functions $\gamma_{\ell m,12}$ that indicate how a given multiple of the SGWB (in a spherical harmonic expansion) affects the cross-correlation between two ground-based detectors. The overlap functions are obtained in terms of two angular ${\hat n}$ integrations, see eq. (\ref{sti-stj}),  with a ${\rm e}^{i \, \Phi_{12}} \equiv {\rm e}^{2 \pi i f {\hat n} \cdot \left( \vec{x}_1 - \vec{x}_2 \right)}$ phase in the integrand. It is instructive to estimate the importance of this phase. The distance between two LIGO sites is about $3000$ km, and the most sensitive LIGO frequencies are of the order of $100$ Hz, so that $\left\vert \Phi \right\vert \simeq 6$ for a correlation between the two LIGO sites. As we show below, the correlators that we compute are instead most sensitive at frequencies of about $7$ Hz (while the sensitivity rapidly worsens at increasing frequencies), corresponding to $\left\vert \Phi \right\vert \simeq 0.18$. This justifies a Taylor expansion of the overlap functions in this phase. In fact, as we show below, for any correlation between specific multipoles of the SGWB either all even of all odd multipoles of $\Phi$ provide a vanishing result. Therefore, if we truncate the Taylor series, the truncated elements correct the result only to ${\rm O} \left( \vert \Phi \vert^2 \right) \simeq 3 \%$. 

Due to the smallness of $\left\vert \Phi \right\vert$, this ET pair would be mostly sensitive to correlators that receive contributions to lowest order in $\Phi$. In particular, we focus our attention to the measurements of the $\ell = 0,\, 2 ,\, 4$ multipoles, for which nonvanishing correlators are obtained already to ${\rm O } \left( \Phi^0 \right)$. As we detail below, when the phase $\Phi$ is altogether disregarded, the overlap functions acquire very simple expressions. In fact, we obtain simple expressions, valid in this regime of small frequency / short separation between the interferometers, that can be applied to any pair of interferometers, each one of arbitrary opening angle between its two arms.~\footnote{These expressons are simpler than those obtained in the literature for LISA \cite{Seto:2004np,Kudoh:2004he,Kudoh:2005as,Taruya:2005yf}, despite the  similar geometry, since the moton of these satellites cannot be described as a rigid rotation about a single fixed axis, as in the case of detectors located on the Earth.} Once applied to the ET pair, these expressions become simple functions of locations of the two sites, and of the two angles $\beta_{1,2}$ that specify the overall orientation of the two ET triangles (for instance, with respect to the north direction at the site). When the different measurements at the ET vertices are combined together, in a way that maximizes the signal to noise ratio (SNR), the dependence on $\beta_{1,2}$ drops, and we obtain an expression for the sensitivity that depends only on the latitude of the two sites, and of the difference between their longitudes. The observation of the other multipoles $\ell$ of the SGWB is suppressed by the smallness of $\vert \Phi \vert$. For these multipoles, we expect a greater sensitivity by correlating ET with other detectors, so to have a longer baseline $\Delta x$ for the observation. We plan to study this in a separate publication. 

The plan of this work is the following. In Section \ref{sec:ET-AET}  we study the noise correlation matrix among different measurements at the same ET site. In Section \ref{sec:signal-smallf} we compute the sensitivity of the ET pair to the SGWB multipoles. Our computations are based on those of ref. 
\cite{Allen:1996gp} that we extend (i) by providing very simple analytical results for the overlap functions in the small frequency regime (namely, at $\Phi = 0$; these expressions are valid for any pair of interferometers), and (ii) by obtaining an immediate final expression for the sensitivity specific to the ET pair. The analytic expressions for the overlap functions are given in  Subsection \ref{subsec:signal}, where we study the expectation value for correlation of the the signal between the two sites. In the following Subsection \ref{subsec:noise} we provide a formal expression for the variance of the measurement due to the instrumental noise in each site. In Subsection \ref{sub:SNR} we then provide a formal expression for the signal-to-noise ratio that combines the various ET measurements of the SWGB. This expression is then evaluated in Subsection \ref{subsec:lm-ET}. In Section \ref{sec:conclusions} we present our conclusions. The paper is completed by four appendices. In Appendix \ref{app:ET-D} we outline how we use the ET sensitivity given in the literature to obtain the variance of the noise at each site needed for our computations. In Appendix \ref{app:e-pc} we derive some useful properties of the GW polarization operators. In Appendix \ref{app:Phi} we show that all odd terms in $\Phi$ do not contribute to the correlators that we have computed, so that our results are corrected only to ${\rm O} \left( \vert \Phi \vert^2 \right)$. Finally, In Appendix \ref{app:coeff} we present the details of the analytic computation of the overlap functions in the small frequency regime.

\section{ET channels and noise diagonalization}
\label{sec:ET-AET}

As mentioned in the Introduction, we assume that we have two identical detectors, each with the same specifications as ET. We denote by $i=1,2$ the $i-$th ET-like detector. The crucial assumption to measure a SGWB below the individual noises of the two detectors is that the noise in one detector is uncorrelated with the noise of the other detector. Each ET-like detector has an equilateral triangular configuration with three Michelson interferometers at its vertices $\alpha = X,\, Y ,\, Z$.  For each interferometer, we measure the difference  $\Delta T$ of the time required by light to complete a return flight across one interferometer arm and that to complete a return flight across the other arm. The measurement is affected by the instrument noise and possibly by a signal,  
\begin{equation}
m_{i\alpha} \left( t \right) = \frac{\Delta T_{i\alpha}}{T_0} = n_{i\alpha} \left( t \right) + s_{i\alpha} \left( t \right) \;\;\;,\;\;\; i =1,\, 2 \;\;\;,\;\;\; \alpha =X,\, Y ,\, Z \;,  
\label{mns}
\end{equation} 
where $T_0$ is the time needed for a  return flight in absence of signal and noise (namely, twice the unperturbed arm length). 

While we assume that the noises in the two detectors are uncorrelated, within each detector the three measurements have correlated noise, due to the fact that every interferometer shares one arm with each of the other two interferometers (of the same detector). Assuming a Gaussian noise with zero mean, we denote its variance in the frequency domain as 
\begin{equation}
\left\langle {\tilde n}_{i\alpha}^* \left( f \right)  {\tilde n}_{j\beta} \left( f' \right) \right\rangle \equiv \frac{\delta_{i j}}{2} \delta_D \left( f - f' \right) N_{\alpha \beta} \left( \left\vert f \right\vert \right) \;, 
\label{variance-ij}
\end{equation}
where the  Kronecker  $\delta_{ij}$ function accounts for the fact that there is no correlation between the noises in the two sites. In the limit in which each detector has an exact equilateral configuration, with identical instruments at the three vertices, the noise correlation matrix at each site is formally of the type
\begin{equation}
N_{\alpha \beta} = \left( \begin{array}{ccc} 
N_d & N_o & N_o \\ 
N_o & N_d & N_o \\ 
N_o & N_o & N_d 
\end{array} \right) \;.  
\label{noise}
\end{equation}
We note that the same correlation matrix is assumed in both ET-like detectors, since the two instruments are assumed to be identical to each other. The noise covariance can be diagonalized by  the three channels 
\begin{equation}
m_{iA} \equiv \frac{2}{3 \sqrt{3}} \left( 2 m_{iX} - m_{iY} - m_{iZ} \right)  \;\;,\;\; 
m_{iE} \equiv \frac{2}{3} \left( m_{iZ} - m_{iY} \right) \;\;,\;\; 
m_{iT} \equiv \left( \frac{2}{3} \right)^{3/2} \left( m_{iX} + m_{iY} + m_{iZ} \right) \;.   
\label{AET-def}
\end{equation} 

These linear combinations were introduced in \cite{Adams:2010vc} for the LISA experiment, that has also an equilateral configuration (the terms $N_d$ and $N_o$ for LISA can be found for example in ref. \cite{Smith:2019wny}). More precisely, we change the normalization of each channel with respect to \cite{Adams:2010vc}, so that the $A-$ and $E-$channels behave as $90^\circ$ degrees interferometers at small frequency, with arm factors $d_{iA,E}^{ab}$ having a standard normalization, see eq. (\ref{dAE-90}).  The redefinition can be written as 
\begin{equation}
m_{iO} = c_{O\alpha} \, m_{i\alpha} \;\;\;\;\;\;,\;\;\;\;\;\; 
c \equiv  \left( \begin{array}{ccc} 
\frac{4}{3 \sqrt{3}}  & - \frac{2}{3 \sqrt{3}} & - \frac{2}{3 \sqrt{3}} \\ 
0 & - \frac{2}{3} & \frac{2}{3} \\ 
\left( \frac{2}{3} \right)^{3/2} &\left( \frac{2}{3} \right)^{3/2} & \left( \frac{2}{3} \right)^{3/2} 
\end{array} \right) \;, 
\end{equation} 
where the index $O$ scans the three channels $A,E,T$, and where we stress that the same combinations $c_{O \alpha}$ are taken in the two detectors. Combining eqs. (\ref{variance-ij}) and (\ref{AET-def}) one obtains 
\begin{eqnarray} 
&& \left\langle {\tilde n}_{iO}^* \left( f \right)  {\tilde n}_{jO'} \left( f' \right) \right\rangle 
= \frac{1}{2} \delta_D \left( f - f' \right) \, \delta_{ij} \, \delta_{OO'} \, N_O \left( \left\vert f \right\vert \right)  \;, 
\label{variance-AET}
\end{eqnarray} 
with 
\begin{equation}
N_A \left( f \right) = N_E \left( f \right) = \frac{8}{9} \left[ N_d \left( f \right) -  N_o \left( f \right) \right] \;\;\;,\;\;\; 
N_T \left( f \right) =  \frac{8}{9} \left[ N_d \left( f \right) + 2  N_o \left( f \right) \right] \;. 
\label{noise-AET}
\end{equation}  
 
We use these channels in this work, as they diagonalize the noise matrix, and this simplifies the computation of the SNR that we perform below. As alerady remarked, and as we explicitly verify in Section \ref{subsec:lm-ET}, the $A-$ and $E-$channels behave as $90^\circ$ interferometers in the small frequency regime (which, as we show below, is the relevant one for the correlators that we are computing). From the literature, we are aware of a computation of the ET sensitivity under the assumption that ET is a single $90^\circ$ interferometer  \cite{Hild:2010id}. Therefore, we use the sensitivity curve of \cite{Hild:2010id} for $N_A \left( f \right) = N_E \left( f \right)$ (see Appendix \ref{app:ET-D} for details). Moreover, in the small frequency regime the $T-$channel vanishes, so we disregard it in our computations.

\section{Measurement of an anisotropic SGWB with two ET-like detectors}
\label{sec:signal-smallf}

We proceed as in ref.  \cite{Allen:1996gp}, that  studied the sensitivity of an Earth-based detector to a directionality-dependent SGWB. Starting from the linear combinations (\ref{AET-def}) of the measurements  (\ref{mns}) at the three vertices, one defines a time-dependent Fourier transform 
\begin{equation}
{\tilde m}_{iO} \left( f ,\, t \right) = \int_{t-\tau/2}^{t+\tau/2} d t' {\rm e}^{-2 \pi i f t'} \, m_{iO} \left( t' \right) \;, 
\label{tmO}
\end{equation}
where the integration is done on a timescale $\tau$ much greater than the inverse of the smallest frequency that we want to study, but sufficiently small that we can disregard the rotation of the Earth in this time. We define ${\tilde s}_{iO} \left( f ,\, t \right) $ and  ${\tilde n}_{iO} \left( f ,\, t \right) $ in an analogous manner. 

One then defines an estimator 
\begin{equation}
{\cal C} \left( t \right) \equiv \sum_{O,O'} \int_{-\infty}^{+\infty} df \, 
{\tilde m}_{1,O}^* \left( f ,\, t \right) {\tilde m}_{2,O'} \left( f ,\, t \right) \, 
{\tilde Q}_{OO'} \left( f \right) \;,  
\label{C-def}
\end{equation}
where the functions ${\tilde Q}_{OO'} \left( f \right) $ are weights (in the sum over channels and frequencies) that will be chosen later to maximize the SNR. 

We assume that the statistical properties of the signal and the noise do not change with time. Then for an anisotropic SGWB, the statistics of the measurement is periodic, with periodicity given by the rotation period $T_e = \frac{2 \pi}{\omega_e}$ of the Earth 
\begin{equation}
{\cal C}  \left( t \right)   = \sum_{m=-\infty}^\infty 
{\cal C}_m \, {\rm e}^{i m \omega_e t} \;\;\;\;,\;\;\;\; 
{\cal C}_m  \equiv  \frac{1}{T} \int_0^{T} dt \, {\rm e}^{-i m \omega_e t} \, 
{\cal C}  \left( t \right) \;, 
\label{C-m}
\end{equation} 
where we take the observation time $T$ to be an integer multiple of one day $T_e$. 

For each coefficient ${\cal C}_m$, we compute the signal-to-noise ratio 
\begin{equation} 
{\rm SNR}_m = \frac{\left\langle {\cal C}_m \right\rangle}{\sqrt{\left\langle {\cal C}_m^2 \right\rangle}} \;. 
\label{SNR}
\end{equation} 

\subsection{Expectation value of the signal } 
\label{subsec:signal}

The expectation value of ${\cal C}_m$ contains only the contribution from the signal 
\begin{equation}
\left\langle {\cal C}_m  \right\rangle =  \frac{1}{T} \int_0^{T} dt \, {\rm e}^{-i m \omega_e t} \,  \sum_{O,O'} \int_{-\infty}^{+\infty} df \, 
\left\langle  {\tilde s}_{1O}^* \left( f ,\, t \right) {\tilde s}_{2O'} \left( f ,\, t \right) \right\rangle 
{\tilde Q}_{OO'} \left( f \right) \;.  
\label{<C>}
\end{equation}

To compute the signal we recall that, to first order in the GW, light starting from $\vec{x}$ at the unperturbed time $t - 2 L$, arriving at $\vec{x} + L \, {\hat l}$, and returning back to $\vec{x}$ completes this flight in the time 
\begin{equation}
T_{\rm return} = 2 L + \frac{{\hat l}^a \, {\hat l}^b}{2} \int_0^L d s \, h_{ab} \left( t - 2 L + s ,\, \vec{x} + s \, {\hat l}  \right) +   \frac{{\hat l}^a \, {\hat l}^b}{2} \int_0^L d s \, h_{ab} \left( t -  L + s ,\, \vec{x} + L \, {\hat l}  - s \, {\hat l}  \right) \;,  
\label{Tret} 
\end{equation}
where we note that $T_0 = 2 L$ is the unperturbed time for a return travel. We work in the regime of low GW frequency / short arm, namely $2 \pi \, f \, L \ll 1$. This applies to existing ground-based interferometers, and to ET (we explicitly verify that this condition applies to our study in  Subsection \ref{sub:SNR}). In this case, we can approximate the GW appearing in (\ref{Tret}) as $h_{ab} \left( t ,\, \vec{x} \right)$, which is constant along the line integral, and therefore 
\begin{equation} 
T_{\rm ret} = 2 L + \frac{{\hat l}^a \, {\hat l}^b}{2} \times 2 \, L \, h_{ab} \left( t ,\, \vec{x} \right) \,.  
\end{equation} 

\begin{figure}[ht!]
\centerline{
\includegraphics[width=0.6\textwidth,angle=0]{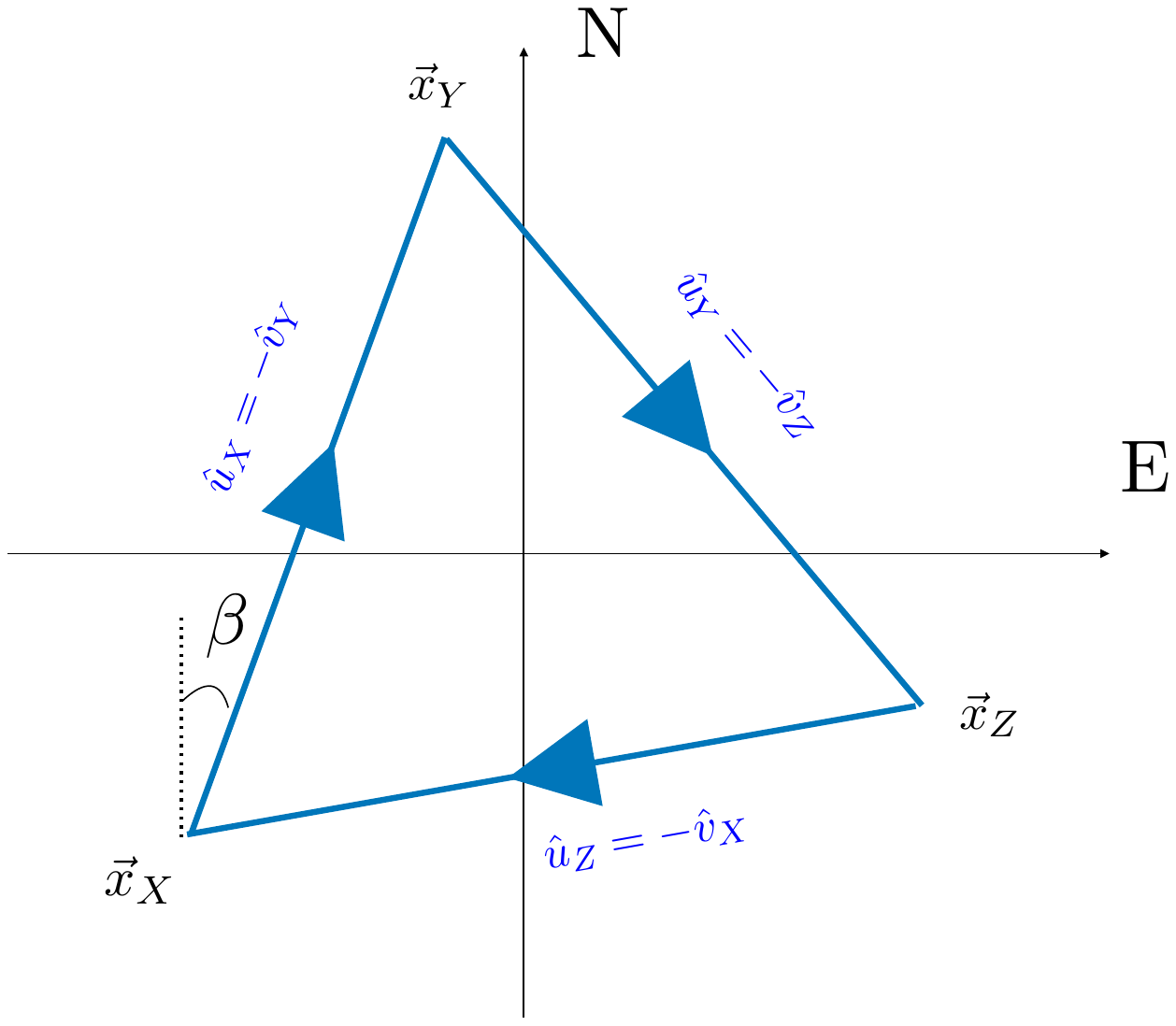}
}
\caption{Our convention for the orientation of the three ET arms, with the labels ``N'' and ``E'' indicating, respectively, the north and east direction at the location of the interferometer, and with $\beta$ being the angle formed by the direction ${\hat u}_X$ of the ``first'' arm and the north direction.  
}
\label{fig:ET2-orientation}
\end{figure}

We denote by $\vec{x}_{i\alpha} \left( t \right)$ the location of the vertex $\alpha$ of the $i-$th detector. We denote by ${\hat u}_{i\alpha} \left( t \right)$ and  ${\hat v}_{i\alpha} \left( t \right)$ the unit vectors in the directions of the two ams that start from this vertex, as indicated in Figure \ref{fig:ET2-orientation} (in the figure, the index $i$ is suppressed, since an identical convention is adopted for the two detectors). We note that these quantities are time-dependent due to the rotation of the Earth about its axis. With this conventions, the signal at the time $t$ at the vertex $\alpha$ of the i-th detector, located at $\vec{x}_{i\alpha}$, is 
\begin{equation}
s_{i\alpha} (t) = d_{i\alpha}^{ab} (t) h_{ab}( t,\vec x_i(t)) \;\;,\;\; 
d_{i\alpha}^{ab} (t) \equiv  \frac{\hat u_{i\alpha}^a(t)\hat u_{i\alpha}^b(t)-\hat v_{i\alpha}^a(t)\hat v_{i\alpha}^b(t)}{2} \;, 
\label{2arms}
\end{equation}

We decompose the GW as 
\begin{equation}
h_{ab}(t,\vec x)=\int_{-\infty}^{\infty}df\int d^2 {\hat n} \, e^{2\pi i f \left( t - {\hat n} \cdot \vec{x} \right)} 
\sum_{s=+,\times} h_s (f,\hat n) \, e^s_{ab} \left( {\hat n} \right) \;, 
\label{eqGWTT}
\end{equation}
where reality is ensured by $h_s^* \left( f, {\hat n} \right) = h_s \left(-f, {\hat n} \right)$, and where the polarization operators $e^s_{ab} \left( {\hat n} \right)$ are discussed in Appendix \ref{app:e-pc}. 

We follow \cite{Allen:1996gp} in assuming an unpolarized anisotropic SGWB, characterized by 
\begin{equation}
\left\langle h^*_s \left( f, {\hat n} \right) h_{r} \left( f' {\hat n}' \right) \right\rangle = \delta_{sr} \delta_D^{(2)} \left( {\hat n} - {\hat n}' \right) \delta_D \left( f-f' \right) H \left( \left\vert f \right\vert \right) P \left( {\hat n} \right) \;, 
\label{eq2points}
\end{equation}
with 
\begin{equation}
P \left( {\hat n} \right) = \sum_{\ell m} p_{\ell m} \, Y_{\ell m} \left( {\hat n} \right) \;,  
\end{equation} 
where in the standard isotropic studies only the monopole is present, with $P(\hat\Omega) = 1$. This angular dependence is formulated in the rest frame of the fixed stars, with the $z-$axis chosen to coincide with the Earth rotation axis. 

We note that the choice made in (\ref{eq2points}) is not the most general one, since it assumes that the frequency and angular dependences are factorized. We also note that only the monopole contributes to the GW energy density, leading to  \cite{Allen:1996gp} 
\begin{equation}
\Omega_{\rm GW} \left( f \right) \equiv \frac{1}{\rho_{\rm critical}} \, \frac{d \rho_{\rm GW}}{d \ln \, f} = 
\frac{32 \pi^3}{3 H_0^2} \, f^3 \, H \left( f \right) \,. 
\label{Omega-H}
\end{equation} 

By using  (\ref{2arms}) and (\ref{eqGWTT}) one finds 
\begin{equation}
{\tilde s}_{i\alpha} \left( f ,\, t \right) = \sum_{s=+,\times} \int d^2 {\hat n} \int_{-\infty}^{+\infty} d f' 
{\rm e}^{-2 \pi i \left( f - f' \right)t} \delta_\tau \left( f - f' \right) 
{\rm e}^{-2 \pi i f' {\hat n} \cdot \vec{x}_{i\alpha} \left( t \right)} 
h_s \left( f' ,\, {\hat n} \right) e_{ab}^s \left( {\hat n} \right) d_{i\alpha}^{ab} \left( t \right) \;, 
\label{sti}
\end{equation} 
where, as we mentioned after (\ref{tmO}), the interferometer location and arms directions can be treated as constant in the time integration of length $\tau$. We have introduced  \cite{Allen:1996gp} 
\begin{equation}
\delta_\tau \left( f \right) \equiv \frac{\sin \left( \pi \tau f \right)}{\pi f} \;\;\;\;,\;\;\;\; 
\lim_{\tau \to \infty} \delta_\tau \left( f \right) = \delta_D \left( f \right) \;. 
\end{equation} 

We then find the correlator 
\begin{eqnarray} 
\left\langle  {\tilde s}_{1\alpha}^* \left( f ,\, t \right) {\tilde s}_{2\beta} \left( f ,\, t \right) \right\rangle &=&  
\sum_{s=+,\times} \int_{-\infty}^{+\infty} d f' \, \delta_\tau^2 \left( f - f' \right) H \left( \left\vert f' \right\vert \right) 
\int d^2 {\hat n} {\rm e}^{2 \pi i f' {\hat n} \cdot \left( \vec{x}_{1\alpha} - \vec{x}_{2\beta} \right)} \nonumber\\ 
&& \quad\quad\quad\quad 
\times \sum_{\ell m} p_{\ell m} Y_{\ell m} \left( {\hat n} \right) \, 
e_{ab}^s \left( {\hat n} \right) e_{cd}^s \left( {\hat n} \right) d_{1\alpha}^{ab} \left( t \right) d_{2\beta}^{cd} \left( t \right) \;. 
\end{eqnarray} 

As the integration time $\tau$ is chosen to be much greater than the inverse of the typical measured frequencies, one of the two $\delta_\tau$ in this expression can be substituted with a Dirac $\delta-$function, while the other one evaluates to the integration time, 
\begin{eqnarray} 
\left\langle  {\tilde s}_{1\alpha}^* \left( f ,\, t \right) {\tilde s}_{2\beta} \left( f ,\, t \right) \right\rangle &=&  \tau 
\sum_{s=+,\times}  H \left( \left\vert f \right\vert \right) 
\int d^2 {\hat n} \, {\rm e}^{2 \pi i f {\hat n} \cdot \left( \vec{x}_{1\alpha} - \vec{x}_{2\beta} \right)} \nonumber\\ 
&& \quad\quad\quad\quad 
\times \sum_{\ell m} p_{\ell m} Y_{\ell m} \left( {\hat n} \right) \, 
e_{ab}^s \left( {\hat n} \right) e_{cd}^s \left( {\hat n} \right) d_{1\alpha}^{ab} \left( t \right) d_{2\beta}^{cd} \left( t \right) \;. 
\label{sti-stj} 
\end{eqnarray} 

In this work, for definiteness, we assume that the two detectors are located at the two sites under consideration for the actual ET detector. The first site is the Sos Enattos mine in Sardinia, at coordinates $40.4^\circ N,\;\; 9.45^\circ E$. The second site is at the Belgium–Netherlands border. For definiteness, we choose the city of Maastricht, at coordinates $50.9^\circ N,\;\; 5.69^\circ E$. Assuming the Earth to be a perfect sphere of radius $R_E = 6,371$ km, the length of the segment connecting these two locations is $d_{\rm ET} \simeq 1,200$ km. For this choice, the argument in the phase of eq. (\ref{sti-stj}) satisfies 
\begin{equation}
\Phi \equiv \left\vert 2 \pi f  {\hat n} \cdot \left( \vec{x}_{1\alpha} - \vec{x}_{2\beta} \right) \right\vert \leq 
2 \pi \, f \, \left\vert \vec{x}_1 - \vec{x}_2 \right\vert \simeq 0.025 \: \frac{f}{\rm Hz} \: \frac{\left\vert \vec{x}_1 - \vec{x}_2 \right\vert}{1,200 \, {\rm km} } \,. 
\label{factor-phase}
\end{equation} 
As we discuss below (see Figure \ref{fig:ET-sensitivity}) this quantity is smaller than one at the frequency to which ET is most sensitive to. As an example, for a scale invariant GW signal ($\Omega_{\rm GW}$ independent of frequency), the sensitivity is strongly peaked at $f \simeq 7 \, {\rm Hz}$. For this value, the product in eq. (\ref{factor-phase}) evaluates to about $0.18$. It is therefore meaningful to evaluate eq. (\ref{sti-stj}) as an expansion series in  $\Phi$. As we show below, the response functions to the various multipoles $p_{\ell m}$ are suppressed by positive powers of $\Phi$, with exception to those to the multipoles $\ell = 0,\, 2 ,\, 4$ and $\left\vert m \right\vert \leq \ell$ \cite{Allen:1996gp}. For this reason, in this work we evaluate only these unsuppressed contributions. As we show in Appendix \ref{app:Phi}, correlators betwwen even (odd) multipoles $\ell$ receive contributions only from even (odd) powers of $\Phi$. Therefore, evaluating these coefficients at $\Phi = 0$ results in a ${\rm O} \left( \Phi^2 \right) \simeq 3 \%$ inaccuracy at the most sensitive frequencies. Taking $\Phi = 0$ allows for a greater simplification of eq. (\ref{sti-stj}). This is a major departure from the study of \cite{Allen:1996gp}, where the much longer distance between the two LIGO interferometers did not allow for this simplification. As we show in this work, in this limit the detector response functions (to be defined shortly) acquire vey simple analytical expressions, which can be employed to determine the SNR, and hence the sensitivity to the anisotropy, almost fully analytically, only up to one numerical integration over frequency. 

Proceeding in this way, we can readily go from the correlation of the signal at the three vertices to the correlator of the signal in the thee channels, and write 
\begin{eqnarray} 
\left\langle  {\tilde s}_{1O}^* \left( f ,\, t \right) {\tilde s}_{2O'} \left( f ,\, t \right) \right\rangle &=&  \tau 
\sum_{s=+,\times}  H \left( \left\vert f \right\vert \right) 
\int d^2 {\hat n} \, \sum_{\ell m} p_{\ell m} Y_{\ell m} \left( {\hat n} \right) \, 
e_{ab}^s \left( {\hat n} \right) e_{cd}^s \left( {\hat n} \right) d_{1O}^{ab} \left( t \right) d_{2O'}^{cd} \left( t \right) \;, 
\nonumber\\ 
\end{eqnarray} 
where $d_{1O}^{ab} \equiv c_{O\alpha} d_{i\alpha}^{ab}$. Evaluating this linear combinations we find 
\begin{equation}
d_{iA}^{ab} = \frac{2}{\sqrt{3}} \, d_{iX}^{ab} \;\;\;\;\;,\;\;\;\;\; 
d_{iE}^{ab} =  - \frac{2}{3}  \left( d_{iX}^{ab}+ 2 d_{iY}^{ab} \right)   \;\;\;\;\;,\;\;\;\;\; 
d_{iT}^{ab} = 0 \;, 
\end{equation}
namely only the two channels A and E are nonvanishing in this limit. Inserting all this in (\ref{<C>}), we obtain 
\begin{eqnarray} 
\left\langle C_m \right\rangle & = &  \frac{\tau}{T} \int_0^T d t \; {\rm e}^{-i m \omega_e t } \sum_{s=+,\times} 
\int_{-\infty}^{+\infty} d f  \, H \left( \vert f \vert  \right) 
\int d^2 {\hat n}  
e_{ab}^s \left( {\hat n} \right) e_{cd}^s \left( {\hat n} \right) \, 
  \sum_{\ell m'} p_{\ell m'} \, Y_{\ell m'} \left( {\hat n} \right) 
\nonumber\\ 
&& \quad\quad\quad\quad  \quad\quad\quad\quad 
\times \, \sum_{O=A,E}  \sum_{O'=A,E}  d^{ab}_{1O} \left( t \right) \, d^{cd}_{2O'} \left( t \right) Q_{OO'} \left( f \right) \;. 
\label{Cm}
\end{eqnarray} 

As we mentioned, the quantities $d^{ab}_{i,A/E} \left( t \right) $ are time dependent because of the rotation of the Earth in the frame of the fixed stars. Denoting by  $d^{ab}_{i,A/E} $ the same quantities in a frame that is fixed with respect to the Earth, 
\begin{equation}
d_{i,A/E}^{ab} \left( t \right) = R_{aa'} \left( t \right) R_{bb'}  \left( t \right) \, d_{i,A/E}^{a'b'} \;, 
\end{equation} 
where $R \left( t \right)$ is a rotation matrix of period $T_e$ around the $z-$axis. We can reabsorb this rotation by changing integration variable ${\hat n} \to R {\hat n}$ in eq. (\ref{Cm}). Using then eq. (\ref{pol-ide2}) we see that the rotation matrix disappears from everywhere apart from the argument of the spherical harmonic, where it produces  $Y_{\ell m'} \left( R {\hat n} \right) = {\rm e}^{i m' \omega_e t}  Y_{\ell m'}  \left(  {\hat n} \right) $. It is then immediate to see that the integration in time then forces $m'=m$, and 
\begin{eqnarray} 
\left\langle C_m \right\rangle & = & \tau \sum_{s=+,\times} 
\int_{-\infty}^{+\infty} d f  \, H \left( \vert f \vert  \right) 
\int d^2 {\hat n}  
e_{ab}^s \left( {\hat n} \right) e_{cd}^s \left( {\hat n} \right) \, 
  \sum_{\ell =\vert m \vert}^\infty p_{\ell m} \, Y_{\ell m} \left( {\hat n} \right) 
\, \sum_{O=A,E}  \sum_{O'=A,E}  d^{ab}_{1O} \, d^{cd}_{2O'}  Q_{OO'} \left( f \right) \;. \nonumber\\ 
\end{eqnarray} 

Following \cite{Allen:1996gp}, we define the overlap functions~\footnote{The factor $\frac{5}{8 \pi}$ is conventional, and it has the purpose of eliminating the overall factor in the last of (\ref{pol-ide1}) in the monopole term.} 
\begin{equation}
\gamma_{\ell m,OO'} \equiv \frac{5}{8 \pi} \int d^2 {\hat n} \, Y_{\ell m} \left( {\hat n} \right) 
 \sum_{s=+,\times} e_{ab}^s \left(  {\hat n} \right) e_{cd}^s \left(  {\hat n} \right) \, d_{1O}^{ab}  \, d_{2O'}^{cd} \;, 
\label{gamma}
\end{equation} 
in terms of which 
\begin{eqnarray} 
\left\langle C_m \right\rangle & = & \frac{8 \pi \, \tau}{5}  
 \int_{-\infty}^{+\infty} d f \, H \left( \vert f \vert \right)   \sum_{\ell = \vert m \vert}^\infty p_{\ell m} \, 
 \sum_{O=A,E}  \sum_{O'=A,E} \gamma_{\ell m,OO'} \,  Q_{OO'} \left( f \right) \;. 
\label{C1-res} 
 \end{eqnarray} 

We note that the functions $\gamma_{\ell m,OO'} $ are frequency-independent, contrary to the analogous functions defined in  \cite{Allen:1996gp}, which do not assume the small frequency regime. We also define the coefficients 
\begin{equation}
\gamma_{\ell m,ab,cd} \equiv \frac{5}{8 \pi} \int d^2 {\hat n} \, Y_{\ell m} \left( {\hat n} \right) 
\sum_{s=+,\times} e_{ab}^s \left(  {\hat n} \right) e_{cd}^s  \left(  {\hat n} \right) 
\;\;\; \Rightarrow \;\;\;  \gamma_{\ell m,OO'}  = \gamma_{\ell m,ab,cd} \times d_{1O}^{ab}  \, d_{2O'}^{cd} \;.  
\label{ga-lm-ab-cd}
\end{equation} 

These coefficients are computed in Appendix \ref{app:coeff}, where, integrating over the two angles, we obtain simple expressions for the $\gamma_{0 m,ab,cd},\, \gamma_{2 m,ab,cd},\, \gamma_{4 m,ab,cd}$ terms: 
\begin{eqnarray}
\gamma_{00,ab,cd} &\cong& \frac{1}{2 \sqrt{\pi}} \left( \delta_{ac} \, \delta_{bd} + \delta_{ad} \, \delta_{bc} \right) \;, 
\nonumber\\ 
\gamma_{20,ab,cd} &\cong& \frac{1}{14} \, \sqrt{\frac{5}{\pi}} \left( 
\delta_{ac} \, A_{bd} +\delta_{ad} \, A_{bc}  + \delta\leftrightarrow A \right) \;, 
\nonumber\\ 
\gamma_{2\pm1,ab,cd} &\cong& 
\frac{3}{14} \sqrt{\frac{5}{6\pi}} \left( 
\delta_{ac} \, B_{bd\pm} + \delta_{ad} \, B_{bc\pm}  +\delta\leftrightarrow B_{\pm} \right) \;, \nonumber\\ 
\gamma_{2\pm2,ab,cd} &\cong&  -  
\frac{3}{14} \sqrt{\frac{5}{6\pi}} \left( 
\delta_{ac} \, C_{bd\pm} + \delta_{ad} \, C_{bc\pm}  +\delta\leftrightarrow C_{\pm} \right) \;, \nonumber\\ 
\gamma_{40,ab,cd} &\cong& \frac{1}{756\sqrt{\pi}} \Bigg[ 
-\delta_{ac} \, \delta_{bd}-\delta_{ad} \, \delta_{bc} 
-5 \left( \delta_{ac} \, A_{bd}+\delta_{ad} \, A_{bc}  +\delta\leftrightarrow A \right) 
\nonumber\\ 
&& \quad\quad \quad\quad +20 \left(A_{ac} \, A_{bd}+A_{ad} \, A_{bc} - \frac{1}{4} A_{ab} \, A_{cd}  \right) \Bigg] \;, \nonumber\\ 
\gamma_{4\pm1,ab,cd} &\cong& \frac{1}{1512} \sqrt{\frac{5}{\pi}} \Bigg[ 2 \left( 
\delta_{ac} \, B_{bd\pm}  + \delta_{ab} \, B_{cd\pm}  +\delta \leftrightarrow B_{\pm} \right) \nonumber\\ 
&& \quad\quad \quad\quad + 7 \left( A_{ac} \, B_{bd\pm} + A_{ad} \, B_{bc\pm}  +  A_{ab} \, B_{cd\pm}  +A \leftrightarrow B_{\pm} \right) \Bigg] \;, \nonumber\\ 
\gamma_{4\pm 2,ab,cd} &\cong& -\frac{1}{504} \sqrt{\frac{5}{2\pi}} \Bigg[
2 \left( \delta_{ac} \, C_{bd\pm} + \delta_{ad} \, C_{bc\pm}  +\delta \leftrightarrow C_{\pm} \right) \nonumber\\ 
&& \quad\quad \quad\quad + 7 \left( A_{ac} \, C_{bd\pm} + A_{ad} \, C_{bc\pm} + A \leftrightarrow C_{\pm} \right) \Bigg] \;, \nonumber\\ 
\gamma_{4\pm 3,ab,cd} & = & - \frac{1}{24} \sqrt{\frac{5}{7\pi}}\left( 
B_{ab\pm} \, C_{cd\pm} + B_\pm \leftrightarrow C_\pm
\right) \;, \nonumber\\ 
\gamma_{4\pm 4,ab,cd} & = & \frac{1}{12} \sqrt{\frac{5}{14\pi}} C_{ab\pm} \, C_{cd\pm}  \;, 
\label{gamma-result}
\end{eqnarray} 
while all the other coefficients vanish. The symbol $\cong$ denotes the fact that we have disregarded terms proportional to $\delta_{ab}$ and to $\delta_{cd}$, as they vanish when contracted with, respectively, the detector coefficients $d_O^{1ab}$ and  $d_{2O'}^{cd}$.  Finally, in eq. (\ref{gamma-result}) we have introduced the matrices 
\begin{align}
A_{cd}=\begin{pmatrix}
1 & 0 & 0 \\
0 & 1 & 0 \\
0 & 0 & -2 \\
\end{pmatrix} \;,\;\;\; 
B_{cd\pm}=\begin{pmatrix}
0 & 0 & \pm 1 \\
0 & 0 & i \\
\pm 1 & i & 0 \\
\end{pmatrix}  \;,\;\;\; 
C_{cd\pm}=\begin{pmatrix}
1 & \pm i & 0 \\
\pm i & -1 & 0 \\
0 & 0 & 0 \\
\end{pmatrix} \, .   
\label{abc} 
\end{align}

The simple analytical expressions (\ref{gamma-result})  are an original result of this work, and they can be used for any pair of detectors (since the geometry of the detectors is encoded in the $d_{1O}^{ab} \, d_{2O'}^{cd}$ term), in the small frequency regime.

\subsection{Variance of the noise } 
\label{subsec:noise}

We evaluate the denominator of eq. (\ref{SNR}) under the assumption of a weak signal, namely assuming that the variance of the signal is negligible with respect to that of the noise. This assumption is valid if one is interested in obtaining the minimum signal that produces an SNR=1. Specifically, we evaluate 
\begin{eqnarray} 
&& \!\!\!\!\!\!\!\! 
\left\langle \left\vert {\cal C}_m \right\vert^2  \right\rangle = \frac{1}{T^2} \int_0^T d t  \int_0^T d t' {\rm e}^{i m \omega_e \left( t - t' \right)} \int_{-\infty}^{+\infty} d f  \int_{-\infty}^{+\infty} d f'  \sum_{O,O',O'',O'''} Q_{OO'}^* \left( f \right)  Q_{O''O'''} \left( f' \right) \nonumber\\ 
&& \quad\quad\quad\quad  \quad\quad\quad\quad \left\langle 
\left[ {\tilde n}_{1O} \left(  f ,\, t \right)  {\tilde n}_{2O'}^* \left( f ,\, t \right)  \right] 
\left[ {\tilde n}_{1O''}^* \left( f' ,\, t' \right)  {\tilde n}_{2O'''} \left( f' ,\, t' \right)  \right] 
\right\rangle \;. 
\label{C2}
\end{eqnarray}

A useful intermediate quantity for this computation is 
\begin{eqnarray} 
\!\!\!\!\!\!\!\! 
\left\langle  {\tilde n}_{iO}^* \left( f ,\, t \right)  {\tilde n}_{jO'} \left( f' ,\, t' \right) \right\rangle = 
\frac{\delta_{ij} \, \delta_{OO'}}{2} \int_{-\infty}^{+\infty} d f_1 \, {\rm e}^{2 \pi i t \left( f - f_1 \right) - 2 \pi i t' \left( f' - f_1 \right) } \delta_\tau \left( f - f_1 \right) \delta_\tau \left( f' - f_1 \right) \, N_O \left( \left\vert f_1 \right\vert \right) \;, 
\nonumber\\ 
\end{eqnarray} 
which follows from combining eq. (\ref{tmO}) for the noise with eq.  (\ref{variance-AET}). It is also useful to recall that reality of $n_O \left( t \right)$ imposes that ${\tilde n}_O \left( f \right) = {\tilde n}_O^* \left( -f \right) $. Inserting this into eq. (\ref{C2}), evaluating the expectation values under the assumption that the noise is gaussian, and integrating over the times, results in 
\begin{eqnarray} 
\left\langle \left\vert {\cal C}_m \right\vert^2  \right\rangle &=& \frac{1}{4 T^2} \int_{-\infty}^{+\infty} d f d f' d f_1 d f_2  \sum_{O,O'=A,E} 
\delta_\tau \left( f - f_1 \right) \delta_\tau \left( f - f_2 \right) 
\delta_\tau \left( f' - f_1 \right) \delta_\tau \left( f' - f_2 \right) \nonumber\\ 
&& \quad\quad\quad\quad 
 \delta_T^2 \left( f_1 - f_2 + \frac{m \omega_e}{2 \pi} \right) \,  N_O \left( \vert f_1 \vert \right)  N_{O'} \left( \vert f_2 \vert \right) 
Q_{OO'}^* \left( f \right) \, Q_{OO'} \left( f' \right)  \;.  
\end{eqnarray} 
The integration time $T$ is much greater than the inverse of the argument of $\delta_T$, so we can treat that term as a Dirac $\delta-$function times $T$. Moreover, we can disregard $\frac{m \, \omega_e}{2 \pi}$ in the argument, as it is much smaller than the frequencies in the ET window. This results in 
\begin{eqnarray} 
\left\langle \left\vert {\cal C}_m \right\vert^2  \right\rangle &=& \frac{1}{4 T} \int_{-\infty}^{+\infty} d f d f' d f_1  \sum_{O,O'=A,E} 
\delta_\tau^2 \left( f - f_1 \right) \delta_\tau^2 \left( f' - f_1 \right)  \nonumber\\ 
&&  N_O \left( \vert f_1 \vert \right)  N_{O'} \left( \vert f_1 \vert \right) 
Q_{OO'}^* \left( f \right) \, Q_{OO'} \left( f' \right) \;. 
\end{eqnarray} 
We treat the $\delta_\tau$ quantities analogously, and we obtain 
\begin{eqnarray} 
\left\langle \left\vert {\cal C}_m \right\vert^2  \right\rangle &=& \frac{\tau^2}{4 T} \int_{-\infty}^{+\infty} d f \sum_{O,O'=A,E}  N_O \left( \vert f \vert \right)  N_{O'} \left( \vert f \vert \right) 
\left\vert Q_{OO'} \left( f \right) \right\vert^2  \;. 
\label{C2-res} 
\end{eqnarray} 

\subsection{SNR computation } 
\label{sub:SNR}

We insert eqs. (\ref{C1-res}) and  (\ref{C2-res}) in the ratio (\ref{SNR}), to obtain 
\begin{equation} 
{\rm SNR}_m = \frac{\left\langle {\cal C}_m \right\rangle}{\sqrt{\left\langle {\cal C}_m^2 \right\rangle}} 
= \frac{\frac{16 \pi \, \sqrt{2T}}{5}  
\int_0^{+\infty} d f \, H \left( f \right)   \sum_{\ell = \vert m \vert}^\infty p_{\ell m} \, 
\sum_{O=A,E}  \sum_{O'=A,E} \gamma_{\ell m,OO'} \,  Q_{OO'} \left( f \right)}{\sqrt{ 
\int_0^{+\infty} d f \, N^2 \left( f \right) \, \sum_{O=A,E}  \sum_{O'=A,E}  \left\vert Q_{OO'} \left( f \right) \right\vert^2 }} \;, 
\end{equation} 
where we have restricted the domain of integration to positive frequencies only, and where we have used the fact that  $N_A \left( f \right) = N_E \left( f \right) \equiv N \left( f \right)$. 

By relabelling 
\begin{eqnarray} 
&& \frac{16 \pi \, \sqrt{2T}}{5}   \, \frac{H \left(  f  \right)}{N \left( f \right)}   \sum_{\ell = \vert m \vert}^\infty p_{\ell m} \,  \gamma_{\ell m,AA}  \equiv {\tilde \gamma}_1 \left( f \right) \;\;,\;\; 
\frac{16 \pi \, \sqrt{2T}}{5}   \, \frac{H \left(  f  \right)}{N \left( f \right)}   \sum_{\ell = \vert m \vert}^\infty p_{\ell m} \,  \gamma_{\ell m,EE}  \equiv {\tilde \gamma}_2 \left( f \right) \;, \nonumber\\ 
&& \frac{16 \pi \, \sqrt{2T}}{5}   \, \frac{H \left(  f  \right)}{N \left( f \right)}   \sum_{\ell = \vert m \vert}^\infty p_{\ell m} \,  \gamma_{\ell m,AE}  \equiv {\tilde \gamma}_3 \left( f \right)  \;\;,\;\; 
\frac{16 \pi \, \sqrt{2T}}{5}   \, \frac{H \left(  f  \right)}{N \left( f \right)}   \sum_{\ell = \vert m \vert}^\infty p_{\ell m} \,  \gamma_{\ell m,EA}  \equiv {\tilde \gamma}_4 \left( f \right) \;, \nonumber\\ 
&& N \left( f \right) \, Q_{AA} \left( f \right)   \equiv {\tilde Q}_1 \left( f \right) \;\;,\;\; 
N \left( f \right) \, Q_{EE} \left( f \right)   \equiv {\tilde Q}_2 \left( f \right) \nonumber\\ 
&& N \left( f \right) \, Q_{AE} \left( f \right)   \equiv {\tilde Q}_3 \left( f \right)  \;\;,\;\; 
N \left( f \right) \, Q_{EA} \left( f \right)   \equiv {\tilde Q}_4 \left( f \right) \;, 
\end{eqnarray} 
we can rewrite 
\begin{equation}
{\rm SNR}_m = \frac{
\int_0^\infty d f \, \sum_{p=1}^4 \, {\tilde \gamma}_p \left( f \right) \, {\tilde Q}_p \left( f \right)
}{
\sqrt{
\int_0^\infty d f \, \sum_{p=1}^4 \, \vert  {\tilde Q}_p \left( f \right) \vert^2
}
} \;, 
\end{equation} 
which is maximized by ${\tilde Q}_p \left( f \right) = c \, {\tilde \gamma}_p^* \left( f \right)$, where $c$ is an arbitrary real constant that can be set to one.  In terms of the original quantities, this gives the final expression for the signal-to-noise ratio 

\begin{eqnarray} 
&& \!\!\!\!\!\!\!\!  \!\!\!\!\!\!\!\!  \!\!\!\!\!\!\!\! 
{\rm SNR}_m =  \frac{3 H_0^2  \, \sqrt{2T}}{10 \, \pi^2} \, \sqrt{ 
\int_0^\infty d f \,  \frac{\Omega_{\rm GW}^2 \left( f \right)}{f^6 \, N^2 \left( f \right)} 
}  \nonumber\\ 
&& \times \left[ 
\left\vert  \sum_{\ell = \vert m \vert}^\infty p_{\ell m} \,  \gamma_{\ell m,AA}  \right\vert^2  + 
\left\vert  \sum_{\ell = \vert m \vert}^\infty p_{\ell m} \,  \gamma_{\ell m,EE}  \right\vert^2   
+ \left\vert  \sum_{\ell = \vert m \vert}^\infty p_{\ell m} \,  \gamma_{\ell m,AE}  \right\vert^2  
+ \left\vert  \sum_{\ell = \vert m \vert}^\infty p_{\ell m} \,  \gamma_{\ell m,EA}  \right\vert^2  
\right]^{1/2} \,, \nonumber\\ 
\label{SNR-result}
\end{eqnarray} 
where also eq. (\ref{Omega-H}) has been used. 

To evaluate eq. (\ref{SNR-result}) we assume a power-law signal in the ET observational window 
\begin{equation}
\Omega_{\rm GW} \left( f \right) = {\bar \Omega}_{\rm GW} \left( \frac{f}{10 \, {\rm Hz} } \right)^\alpha \;, 
\end{equation}
where ${\bar \Omega}_{\rm GW}$ is the fractional energy density at the pivot scale of $10 \, {\rm Hz}$. 
Typical values considered for the spectral index are $\alpha=0$, as for a cosmological inflationary signal (characterized by nearly scale-invariance) and $\alpha = 2/3$, as expected for the stochastic background due to black hole-black hole and black hole-neutron star binary system inspirals \cite{Abbott:2017xzg}. We can then rewrite eq. (\ref{SNR-result}) as 
\begin{eqnarray}
{\rm SNR}_m &=& \sqrt{\frac{T}{1 \, {\rm year}}} \, {\cal F}_\alpha 
\times 
\Bigg[ 
\left\vert  \sum_{\ell = \vert m \vert}^\infty  {\bar \Omega}_{\rm GW} \, p_{\ell m} \,  \gamma_{\ell m,AA}  \right\vert^2  + 
\left\vert  \sum_{\ell = \vert m \vert}^\infty  {\bar \Omega}_{\rm GW} \, p_{\ell m} \,  \gamma_{\ell m,EE}  \right\vert^2    \nonumber\\ 
&& \quad\quad\quad\quad + 
\left\vert  \sum_{\ell = \vert m \vert}^\infty  {\bar \Omega}_{\rm GW} \, p_{\ell m} \,  \gamma_{\ell m,AE}  \right\vert^2   
+  \left\vert  \sum_{\ell = \vert m \vert}^\infty  {\bar \Omega}_{\rm GW} \, p_{\ell m} \,  \gamma_{\ell m,EA}  \right\vert^2  
\Bigg]^{1/2} \;. \nonumber\\  
\label{SNR2}
\end{eqnarray} 
where we have normalized the total observation time to one year, and where we have defined the dimensionless factor 
 \begin{equation}
{\cal F}_\alpha \equiv \sqrt{ \frac{1 \, {\rm year}}{\left( 1 \, {\rm Hz} \right)^3 } \, \frac{9 H_0^4}{2\pi^4} \, \int_{f_{\rm min}}^{f_{\rm max}} \frac{df}{1 \, {\rm Hz} } \, \frac{ 10^{-2 \alpha} \left( \frac{f}{1 \, {\rm Hz} } \right)^{2\alpha-6}}{S_n^2 \left( f \right) \, {\rm Hz}^2} } 
\simeq \sqrt{ \int_{\rm Hz}^{10^4 \, {\rm Hz}} \frac{df}{1 \, {\rm Hz} } \, \frac{ 3.2 \times 10^{-65-2 \alpha} \,\left( \frac{f}{1 \, {\rm Hz} } \right)^{2\alpha-6}}{S_n^2 \left( f \right) \, {\rm Hz}^2} }  \,, 
\label{calF} 
\end{equation} 
in which eq. (\ref{NSn}) has been used.  In the evaluation we have taken a year of $365.25$ days, the value of the current Hubble rate $H_0 \simeq 67 \, {\rm km} \, {\rm s}^{-1} \, {\rm Mpc}^{-1}$ indicated by Planck \cite{Aghanim:2018eyx}, and the minimum and maximum ET frequencies given in \cite{ET-Punturo-ETD}.

\begin{figure}[ht!]
\centerline{
\includegraphics[width=0.45\textwidth,angle=0]{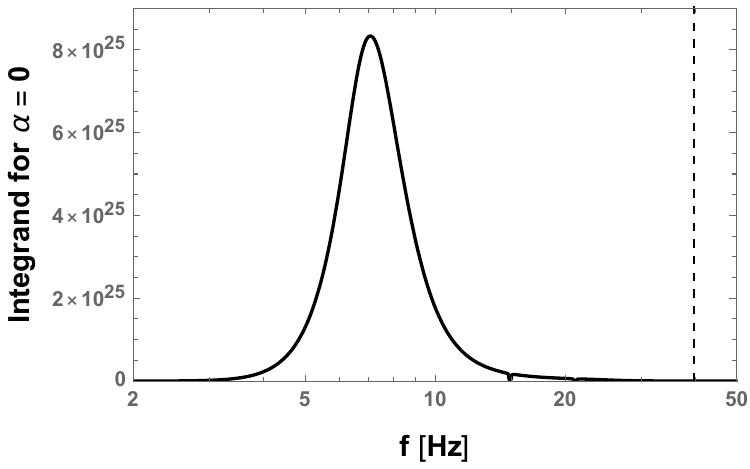}
\includegraphics[width=0.45\textwidth,angle=0]{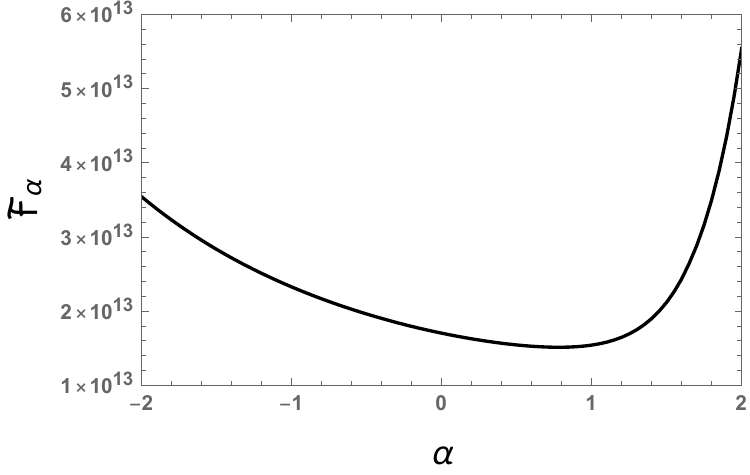}
}
\caption{Left panel: integrand of the quantity ${\cal F}_\alpha$ introduced in eqs. (\ref{SNR2}) and (\ref{calF}), for a scale-invariant $\Omega_{\rm GW}$. The vertical dashed line is the threshold (\ref{condition2}) for the low frequency / short separation condition, in the hypothesis in which two ET-like instruments are placed at the sites currently under consideration for ET. We see that the ET sensitivity is completely dominated by frequencies that satisfy this condition. Right panel: Value of ${\cal F}_\alpha$ for $\alpha$ ranging between $-2$ and $2$. 
}
\label{fig:ET-sensitivity}
\end{figure}

In the left panel of Figure \ref{fig:ET-sensitivity} we show the integrand of eq. (\ref{calF}) for the choice of $\alpha = 0$. As discussed after eq. (\ref{factor-phase}), our results are valid for frequencies 
\begin{equation}
f \ll \frac{1}{2 \pi \, d_{\rm ET}} \simeq 40 \, {\rm Hz} \frac{1,200 \, {\rm km}}{d_{ET} } \;, 
\label{condition2} 
\end{equation} 
where $d_{ET} = 1,200 \, {\rm Km}$ is the distance of the segment connecting the two sites under consideration for ET. We see that the sensitivity is completely dominated by frequencies that satisfy the condition (\ref{condition2}). As already remarked, disregarding the phase in eq. (\ref{sti-stj}) amounts in a ${\rm O} \left( \frac{f}{40 \, {\rm Hz}} \right)^2$ mistake, that, evaluated at the peak frequency visible in the figure, is about $0.03$. In the right panel of Figure \ref{fig:ET-sensitivity} we plot the value of ${\cal F}_\alpha$  for $\alpha$ ranging between $-2$ and $2$. The two cases $\alpha=0,2/3$ mentioned above correspond, respectively, to ${\cal F}_0 \simeq 1.7 \times 10^{13}$ and ${\cal F}_{2/3} \simeq 1.5 \times 10^{13}$.

\subsection{Sensitivity of the ET pair to multipoles of the SGWB } 
\label{subsec:lm-ET}

In this subsection we first assume that only one multiple $p_{\ell m}$ dominates the SGWB, so that 
\begin{eqnarray}
&& \!\!\!\!\!\!\!\! 
{\rm SNR}_m =  {\cal F}_\alpha \, \sqrt{\frac{T}{1 \, {\rm year}}} \,  {\bar \Omega}_{\rm GW} \left\vert  p_{\ell m} \right\vert \left[ 
\left\vert      \gamma_{\ell m,AA}  \right\vert^2  + 
\left\vert     \gamma_{\ell m,EE}  \right\vert^2   
+  \left\vert    \gamma_{\ell m,AE}  \right\vert^2  +  \left\vert    \gamma_{\ell m,EA}  \right\vert^2  
\right]^{1/2} \;. 
\end{eqnarray} 
The threshold amplitude to give ${\rm SNR} =1$ is therefore 
\begin{equation} 
{\bar \Omega}_{\rm GW} \left\vert  p_{\ell m} \right\vert \Bigg\vert_{\rm threshold} = 
\sqrt{\frac{1 \, {\rm year}}{T}} \, 
\frac{1}{ {\cal F}_\alpha \;\; \gamma_{\ell m,{\rm combined}} } \;, 
\end{equation} 
where~\footnote{We note that $\gamma_{\ell, -m,OO'} = \left( - 1 \right)^m  \gamma_{\ell, -m,OO'}^*$, due to the property of the spherical harmonics in eq. (\ref{gamma}). Therefore the value of $\gamma_{\ell m,{\rm combined}}$ does not depend on the sign of $m$.}
\begin{equation} 
\gamma_{\ell m,{\rm combined}} \equiv 
\left[ 
\left\vert      \gamma_{\ell m,AA}  \right\vert^2  + 
\left\vert     \gamma_{\ell m,EE}  \right\vert^2   
+  \left\vert    \gamma_{\ell m,AE}  \right\vert^2  
+  \left\vert    \gamma_{\ell m,EA}  \right\vert^2  
\right]^{1/2} \;. 
\label{ga-comb}
\end{equation} 
As expected, the threshold value decreases as the inverse of the square root of the observation time. For definiteness, we fix $T$ to one year in the following computations. 

To evaluate the threshold value, we need to compute the overlap function elements according to eq. (\ref{ga-lm-ab-cd}). In eq. (\ref{gamma-result})  we provided simple analytic results for the detector-independent $\gamma_{\ell m,ab,cd}$ coefficients. We now need to determine the detector-dependent  elements $d_{iA}^{ab}$ and  $d_{iE}^{ab}$, which encode the orientation of the arms of the ET pair. 
 
From the definition (\ref{gamma}), and from the property (\ref{pol-ide2}), we see that $\gamma_{\ell m, OO '}$ transforms as $Y_{\ell m}$ under a rotation. Namely, 
\begin{eqnarray} 
\gamma_{\ell m,RORO'} &=& \frac{5}{8 \pi} \int d^2 {\hat n} \, Y_{\ell m} \left( {\hat n} \right) 
\sum_{s=+,\times} e_{ab}^s \left(  {\hat n} \right) e_{cd}^s  \left(  {\hat n} \right) 
\, R^a_{a'}  R^b_{b'} d_{1O}^{a'b'} 
\, R^c_{c'}  R^d_{d'} d_{2O'}^{c'd'} \nonumber\\ 
&=& \frac{5}{8 \pi} \int d^2 {\hat n} \, Y_{\ell m} \left( R {\hat n} \right) 
\sum_{s=+,\times} e_{ab}^s \left(  {\hat n} \right) e_{cd}^s  \left(  {\hat n} \right) 
d_{1O}^{ab}  d_{2O'}^{cd} \;. 
\label{gamma-rot}
\end{eqnarray} 
 
Therefore, for a rotation of angle $\phi$ about the  $z-$axis, $\gamma_{\ell m,OO'} \to {\rm e}^{i m \phi} \, \gamma_{\ell m,OO'}$. It follows that $\left\vert \gamma_{\ell m,OO'} \right\vert^2$ is invariant under such rotation. With our choice of frame, a rotation about the $z-$axis connects two locations on Earth that have the same latitude, and different longitude. It follows that the sensitivity to the various multipoles does not depend on the individual longitudes of the two sites, but only on their difference. 

We specify the position and orientation of the two ET-like detectors as follow. We denote the latitude and longitude of each site with $\theta_i$ and $\phi_i$, respectively (we recall that $i=1,2$ denote the first or the second detector). Following standard convention, the north pole is at latitude $\theta =0$, while the equator is at latitude $\theta = \pi/2$. Let us consider a Cartesian system centred at the center of the Earth (assumed to be a perfect sphere), with the $x-$ axis pointing toward the location of $0$ longitude on the equator, with the $y-$ axis pointing toward the location of $\pi/2$ longitude on the equator, and with the $z-$axis pointing toward the north pole. In this coordinate system, at the $\left\{ \theta_i ,\, \phi_i \right\}$ location, the north direction is given by 
\begin{equation}
{\hat v}_{i,{\rm north}}  = 
\left\{ - \sin \theta_i \, \cos \phi_i ,\, - \sin \theta_i \, \sin \phi_i ,\, \cos \theta_i \right\} \;, 
\end{equation} 
while the east direction is given by 
\begin{equation} 
{\hat v}_{i,{\rm east}} = 
\left\{ - \sin \phi_i ,\,  \cos \phi_i ,\, 0 \right\} \;. 
\end{equation} 
we note that these vectors are a basis for the tangent space to the Earth surface at the $\left\{ \theta_i ,\, \phi_i \right\}$ location, and they can therefore be employed to specify the directions of the arms of the  detectors. In our computations we specify the orientation by the angle $\beta_i$  that the arm direction ${\hat u}_{iX}$ forms with the north direction, 
\begin{equation}
{\hat u}_{iX} = \cos \left( \beta_i \right) \, {\hat v}_{\rm north} +  \sin \left( \beta_i \right) \, {\hat v}_{\rm east} \;, 
\label{orientation-plane} 
\end{equation} 
and then following the conventions indicated in Figure \ref{fig:ET2-orientation}. 

An explicit evaluation of the $d_{iA}^{ab}$ and  $d_{iE}^{ab}$ coefficients show that the $A-$ and the $E-$ channels can be understood as two $90^\circ$ interferometers that are shifted by $45^\circ$ with respect to each other. To see this, we can imagine placing one interferometer to the north pole ($\theta = 0$), arriving to it from the meridian that joins the equator along the negative $y-$axis ($\phi = - \frac{\pi}{2}$). In this way the ${\hat v}_{\rm east}$ and the ${\hat v}_{\rm north}$ directions are unit vectors oriented, respectively, along the $x-$ and $y-$ axis (as a conventional $2d$ Cartesian system in Figure \ref{fig:ET2-orientation}.) 
One then finds that 
\begin{eqnarray} 
d_A^{ab} \left( \beta \right) =   d_{90^\circ}^{ab} \left( \frac{7 \pi}{12} - \beta \right) \;\;,\;\; 
d_E^{ab} \left( \beta \right) = d_A^{ab} \left( \beta + \frac{\pi}{4} \right)  \;\;, 
\label{dAE-90}
\end{eqnarray} 
where $d_{90^\circ}^{ab} \left( \alpha \right)$ is the element in eq. (\ref{2arms}) resulting from a $90^\circ$ interferometer with the ${\hat u}$ arm (resp., ${\hat v}$ arm) oriented at a counter-clockwise angle $\alpha$ 
(resp., $\alpha + 90^\circ$) with respect to the $x-$axis. 
 
\begin{figure}[ht!]
\centerline{
\includegraphics[width=0.5\textwidth,angle=0]{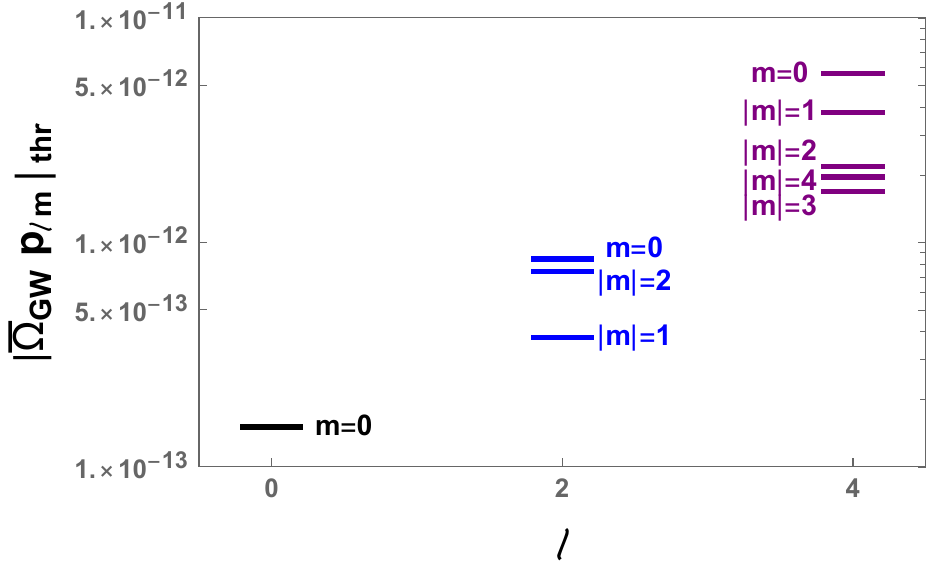}
}
\caption{
Sensitivity to the monopole and to the various quadrupole and hexadecapole  moments. 
The solid lines shows the amplitude that the multipoles must have to produce SNR = 1 in one year of observation ot two ET-like detectors placed at the two sites currently under consideration for ET.  A scale invariant $\Omega_{\rm GW}$ is assumed ($\alpha = 0$). 
}
\label{fig:plm-sensitivity}
\end{figure}

As a consequence of the first of (\ref{dAE-90}), $d_{A}^{ab} \left( \beta + \frac{\pi}{2} \right) = - d_A^{ab} \left( \beta  \right)$. Combining this with the second of (\ref{dAE-90}) results in $d_{A}^{ab} \left( \beta \right) = - d_{E}^{ab} \left( \beta + \frac{\pi}{4} \right)$.  Therefore, the combination (\ref{ga-comb}) is at least invariant under the independent variations $\beta_1 \to \beta_1+  \frac{\pi}{4}$ and / or $\beta_2 \to \beta_2+  \frac{\pi}{4}$. One also expects that any physical result, such as the optimal SNR, should be invariant under a different labeling of the three arms. Therefore  (\ref{ga-comb}) should also be invariant under the independent  variations $\beta_1 \to \beta_1 + \frac{2 \pi}{3}$ and / or $\beta_2 \to \beta_2 + \frac{2 \pi}{3}$.  Combining these two periodicities, we see that  (\ref{ga-comb}) should at least be invariant for  $\beta_i \to \beta_i + \frac{\pi}{12}$. In fact, an explicit evaluation of these coefficients shows that the combination  (\ref{ga-comb}) is independent of $\beta_i$, and it therefore depends only on the latitude of the detectors, $\theta_1$ and $\theta_2$, and on the difference of their latitudes, $\phi_2 - \phi_1$. Evaluating the coefficients between the two sites currently under ccnsideration for ET produces the sensitivities shown in Figure \ref{fig:plm-sensitivity}. 

Finally, to give a measure of the sensitivity to a given multipole number $\ell$, we consider the case of a statistically isotropically signal, 
\begin{equation}
\left\langle a_{\ell m}  a_{\ell' m'}^*  \right\rangle = C_\ell \, \delta_{\ell \ell'} \, \delta_{m m'} \;, 
\end{equation} 
with a single $C_\ell$ dominating over the other ones. We obtain the following expectation from the $m=0,2,4$ measurements 
\begin{eqnarray} 
\left\langle \sum_m {\rm SNR}_m \right\rangle &=& \sqrt{ 
\frac{T}{1 \, {\rm year}} \, {\cal F}_\alpha^2 \,  {\bar \Omega}_{\rm GW}^2 \, \sum_m 
\left\langle  p_{\ell m}  p_{\ell m}^* \right\rangle  
\sum_{O=A,E} \sum_{O'=A,E} 
\gamma_{\ell m,OO'}  \,  \gamma_{\ell m,OO'}^* 
}  \nonumber\\  
&=& \sqrt{\frac{T}{1 \, {\rm year}} } \, {\cal F}_\alpha \, C_{\ell}^{1/2} \,  {\bar \Omega}_{\rm GW}
\left[ \sum_m \left( \left\vert \gamma_{\ell m,AA} \right\vert^2 +  \left\vert \gamma_{\ell m,AE} \right\vert^2 +  \left\vert \gamma_{\ell m,EA} \right\vert^2 +  \left\vert \gamma_{\ell m,EE} \right\vert^2 \right) \right]^{1/2} \nonumber\\ 
&=& \sqrt{\frac{T}{1 \, {\rm year}} } \, {\cal F}_\alpha \, C_{\ell}^{1/2} \,  {\bar \Omega}_{\rm GW} 
\left[ \sum_m \gamma_{\ell m,{\rm combined}}^2 \right]^{1/2} \nonumber\\ 
&\equiv& \sqrt{\frac{T}{1 \, {\rm year}} } \, {\cal F}_\alpha \, C_{\ell}^{1/2} \,  {\bar \Omega}_{\rm GW} 
\, \gamma_{\ell,{\rm tot}} 
\end{eqnarray} 
and the sensitivity to $C_\ell$ is then obtained from the sum over all $m$'s
\begin{equation}
{\bar \Omega}_{\rm GW} \, C_\ell^{1/2} \Bigg\vert_{\rm threshold, 1 year} = 
\frac{1}{ {\cal F}_\alpha \;\; \left[ \sum_{m=-\ell}^\ell \gamma_{\ell m,{\rm combined}}^2 \right]^{1/2}} 
\label{thr-iso}
\end{equation} 

\begin{figure}[ht!]
\centerline{
\includegraphics[width=0.5\textwidth,angle=0]{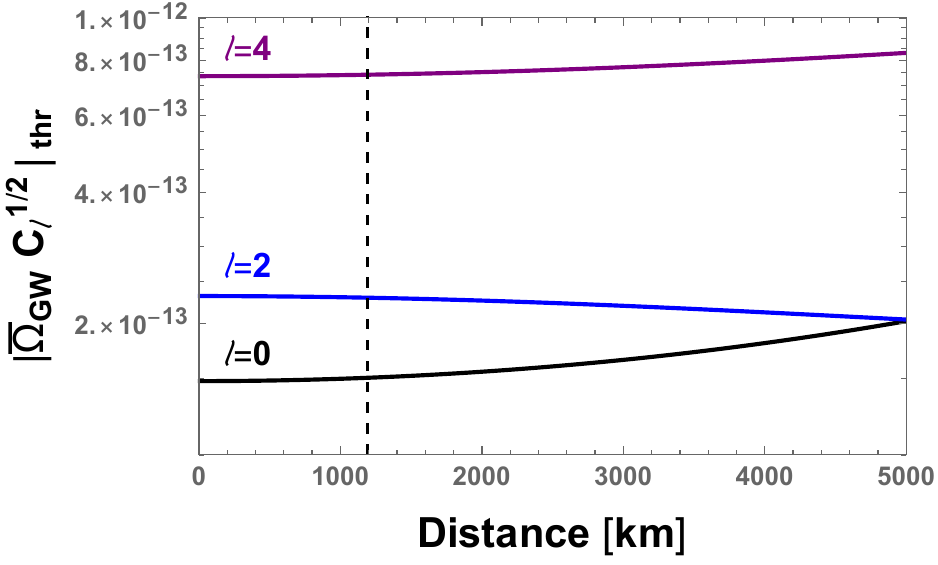}
}
\caption{
Sensitivity to a statistically invariant monopole, quadruple, and hexadecapole multipole of the SGWB. The lines show the amplitude that the multipoles must have to produce SNR = 1 in one year of observation. The horizontal axis is the distance between the two sites. The vertical dashed line corresponds to the distance between the two sites under consideration for ET.  A scale invariant $\Omega_{\rm GW}$ is assumed ($\alpha$ = 0).
}
\label{fig:Cl-sensitivity}
\end{figure}

Given what we proved in eq. (\ref{gamma-rot}), summing over $m$ results in a quantity that is invariant under rotations \cite{Kudoh:2004he}. Therefore, the sensitivity (\ref{thr-iso}) only depends on the opening angle $\psi$ formed by the two radial vectors that, starting from the  center of the Earth, point in the directions of the two sites. Or, equivalently, on the distance $R_E \, \psi$ between the two sites.~\footnote{This distance is the length of the arc on the Earth surface between the two sites. The angle between the two sites under considerations for ET is $\psi \simeq 0.19$. For this angle, the length of the arc is only about 0.15 \% greater than the length of the segment hoining the two sites, that is the distance that should be used in the condition (\ref{condition2}).} An explicit evaluation gives 
\begin{eqnarray}
\gamma_{0,{\rm tot}}^2 &=& \frac{1 + 6 \cos^2 \psi + \cos^4 \psi}{16 \pi} \;\;, \nonumber\\ 
\gamma_{2,{\rm tot}}^2 &=& \frac{5 \left[ 13 - 6 \cos^2 \psi + \cos^4 \psi \right] }{196 \pi} \;\;, \nonumber\\ 
\gamma_{4,{\rm tot}}^2 &=& \frac{321 + 246 \cos^2 \psi + \cos^4 \psi}{28224 \pi} \;\;. 
\end{eqnarray} 
We show the corresponding threshold values in Figure \ref{fig:Cl-sensitivity}. 

\section{Conclusions}
\label{sec:conclusions} 

In this work we studied the sensitivity of an ET pair, under the hypothetical assumption that two instruments are built at the two sites under current consideration for ET, to various multipoles of the SGWB. This pair would operate in the small frequency / short separation regime, namely the distance $d \simeq 1,200 \, {\rm km}$ between the sites is much smaller than the inverse of the most sensitive frequency $f \simeq 7 \, {\rm Hz}$. The sensitivity to the various multipoles $p_{\ell m}$ of the SGWB is suppressed by $f \, d \ll1$, with the exception of the $\ell = 0,2,4$ multipoles. For this reason, we concentrated our study to these unsuppressed multipoles. 

Our computation is based on that of ref. \cite{Allen:1996gp}, that provided the formalism to study the response functions to the anisotropic SGWB from ground-based detectors, that have a well defined scanning pattern related to the daily rotation of the Earth. We extended their results in several ways. Firstly, we provided very simple analytic expressions for the overlap functions in the $f \, d \ll1$ regime. In this limit, a phase in the detector overlap functions to the multipoles of the SGWB, that depends on the frequency, and on the scalar product between the GW arrival direction and the displacement between the detector interferometers, is negligible. This allows to obtain very simple coefficients, reported in eq. (\ref{gamma-result}) that, once contracted with the arm directions, provide the response functions for any given pair of detectors. The coefficients  (\ref{gamma-result})  are independent of frequency and of the geometry of the detectors, and therefore they can be used to compute the overlap function of any interferometer (if one is interested in a self-correlation) or to any pair of interferometers (if one is interested in a cross correlation) in the  small frequency / short separation regime. In some applications, such as the one studied here, this regime applies to the most sensitive part of the frequency observational window. 

Secondly, we specified this study to ET geometries. This instrument will be composed of three nested detectors, located at the vertices of an equilateral triangle. Each detector therefore shares one arm direction with each of the other two detectors, generating a cross correlated noise. We perform computations for the so called A,E,T channels, which are essentially linear combinations of the measurements taken at the three vertices. These combinations were introduced in \cite{Adams:2010vc} to diagonalize the noise correlation matrix in the LISA experiment (that also has three detectors on the vertices of an equilateral triangle). Eq. (\ref{SNR2}) shows how to combine the various measurements taken in these channels so to maximize the sensitivity to the multipoles of the SGWB. We then studied the sensitivity to the various $\ell =0,\, 2 ,\, 4$ (and varying $m$) multipoles, namely the amount of signal that each multipole (assumed to dominate the SGWB) needs to have to be detected at the ET pair. We found that the sensitivity is only function of the latitude of the two ET sites and on the difference of their longitude, but not on the orientation of the two ET triangles. 

The present work can be extended in a number of directions, for example by computing the response functions beyond the small frequency / short separation regime, so to quantify the sensitivity to other multipoles of the SWGB, or by computing the overlap between ET and other ground-based detectors (in particular, the 3G Cosmic Explorer \cite{Reitze:2019iox}), which will provide a longer beaseline for the measurements of the multipoles with $\ell \neq 0,\, 2 ,\, 4$.  We hope to come back to these analyses in a separate publication.

\vskip.25cm
\section*{Acknowledgements} 
We thank Carlo R. Contaldi,  Lorenzo Sorbo, and Gianmassimo Tasinato for discussions on the measurement of the anisotropies of the SWGB. We thank Vuk Mandic, Angelo Ricciardone, and an anonymous referee for discussions and for very useful comments on the manuscript. We particularly thank Robert R. Caldwell for sharing some unpublished notes on the detection of the anisotropic SGWB with LISA.

\vskip.25cm

\appendix

\section{Sensitivity of ET}
\label{app:ET-D}

In this Appendix we discuss how to map the ET sensitivity curve given in the literature to the noise term $N_A \left( f \right) = N_E \left( f \right)$ used in the main text. We stress that we are computing the correlation (\ref{C-def}) of the measurements at two different sites, which, by assumption, have uncorrelated noise. The noise at each individual site affects the variance of the measurements  (entering at the denominator of the signal-to-noise ratio), which is the effect that we are computing here. 

As we show in the main text, the $A-$ and $E-$ channels behave as two $90^\circ$ interferometers, with identical and uncorrelated noise, see eqs. (\ref{variance-AET}) and (\ref{noise-AET}). To our knowledge, the ``ET-D'' sensitivity curve of \cite{Hild:2010id} is the state of the art result for ET \cite{Maggiore:2019uih}. This curve has been computed precisely for a $90^\circ$ interferometer, and it can be obtained from \cite{ET-Punturo-ETD}. The sensitivity curve is plotted at the Website \cite{GWplotter} as the ET Power Spectral Density $S_n \left( f \right)$. According to the companion paper \cite{Moore:2014lga} this quantity is the error on the PSD of the signal $S_h \left( f \right)$. The latter is related to the GW fractional energy density by 
\begin{equation} 
S_h \left( f \right) = \frac{3 H_0^2 \, \Omega_{\rm GW} \left( f \right)}{2 \pi^2 f^3} \,. 
\label{Omega-SH} 
\end{equation} 

To relate $N_A = N_E \equiv N$ to $S_n$ we perform the computation of the SNR for an isotropic SGWB  measured by a single $90^\circ$ angle interferometer (we are not claiming that this measurement is actually feasible, as one would require an extremely good knowledge of the noise function, which can never be measured in absence of the signal; we stress that this computation is done only to estimate the quantity $N$ to be used in the noise variance at each site from the ET Power Spectral Density given in the literature. In absence of noise correlation between the two sites, an imperfect knowledge of the variance of the noise at each sites can affect the precise SNR estimate, but not the fact that the SNR grows as $\sqrt{T}$). For this purpose, we consider the estimator 
\begin{equation}
{\tilde C} \equiv \frac{1}{T} \int_0^T d t \int_{-\infty}^{+\infty} d f 
\left[ {\tilde m}^* \left( f ,\, t \right) {\tilde m} \left( f ,\, t \right) 
- \left\langle  {\tilde n} \left( f ,\, t \right) {\tilde n} \left( f ,\, t \right) \right\rangle \right] 
{\tilde Q} \left( f \right) \;, 
\end{equation}
where $m= s+n$ denotes the fractional time difference measured at this interferometer, where the signal is 
still given by eq. (\ref{sti}), with $P \left( {\hat n} \right) = 1$ in  eq. (\ref{eq2points}), and the noise satisfies 
\begin{eqnarray} 
\left\langle {\tilde n}^* \left( f \right)  {\tilde n} \left( f' \right) \right\rangle 
= \frac{1}{2} \delta_D \left( f - f' \right) \,  N \left( \left\vert f \right\vert \right)  \;, 
\end{eqnarray} 

The SNR computation proceeds as in the main text, leading to the optimal SNR 
\begin{equation}
{\rm SNR} = \frac{16 \pi \sqrt{T}}{5} \sqrt{\int_0^\infty d f \, \left( \frac{H \left( f \right)}{N \left( f \right)} \right)^2 } =  \sqrt{T \, \int_0^\infty d f \, \left( \frac{S_h \left( f \right)}{5 \, N \left( f \right)} \right)^2 } 
 \;. 
\end{equation} 
where eqs. (\ref{Omega-H}) and (\ref{Omega-SH}) have been used in the last step. This leads us to the identification 
\begin{equation}
5 \, N \left( f \right) = S_n \left( f \right) \;. 
\label{NSn}
\end{equation} 
In summary, the function $N \left( f \right)$ to be used in the main text is $\frac{1}{5}$ times the sensitivity curve given in \cite{Hild:2010id}.

\section{Polarization operators}
\label{app:e-pc}

The GW polarization operators are customarily defined as 
\begin{align} 
e_{ab}^+ \left( {\hat n} \right) \equiv m_am_b-n_an_b \;\;,\;\; 
e_{ab}^\times \left( {\hat n} \right) \equiv m_an_b+n_am_b \;, 
\label{matrPolarizz}
\end{align} 
with 
\begin{eqnarray} 
m \left( {\hat n} \right) &\equiv& \frac{{\hat n} \times {\hat e}_z}{\vert {\hat n} \times {\hat e}_z \vert } = 
\left( \sin \phi ,\, - \cos \phi ,\, 0 \right)  \;, \nonumber\\ 
n \left( {\hat n} \right) &\equiv& {\hat n} \times  m =  \left( \cos \theta \, \cos \phi ,\,  \cos \theta \, \sin \phi ,\, - \sin \theta \right) \;, 
\end{eqnarray} 
where ${\hat e}_z$ is the unit vector along the third axis, while 
${\hat n} = \left( \sin \theta \, \cos \phi ,\,  \sin \theta \, \sin \phi ,\, \cos \theta \right)$ is the direction of propagation of the GW. Direct evaluations shows that 
\begin{eqnarray} 
e_{ab}^r \left( {\hat n} \right) \, e_{ab}^s \left( {\hat n} \right) &=& 2 \, \delta^{rs} \;, \nonumber\\ 
\sum_{s=+,\times} e_{ab}^s \left( {\hat n} \right)  e_{cd}^s \left( {\hat n} \right) &=& 
{\bar Q}_{ac} \, {\bar Q}_{bd} + {\bar Q}_{ad} \, {\bar Q}_{bc} - {\bar Q}_{ab} \, {\bar Q}_{cd} \;\;\;,\;\;\; 
{\bar Q}_{ab} \equiv \delta_{ab} - {\hat n}_a \, {\hat n}_b \;, \nonumber\\ 
\sum_{s=+,\times} \int d^2 {\hat n} \, e_{ab}^s \left( {\hat n} \right)  e_{cd}^s \left( {\hat n} \right) &=& 
\frac{8 \pi}{5} \left( \delta_{ac} \delta_{bd} + \delta_{ad} \delta_{bc} - \frac{2}{3} \delta_{ab} \delta_{cd} \right) \;. 
\label{pol-ide1}
\end{eqnarray} 

We also introduce the helicity operators 
\begin{equation}
{\tilde e}_{ab,R} \equiv \frac{e_{ab}^+ + i \, e_{ab}^\times}{\sqrt{2}} \equiv {\tilde e}_{ab,1} \;\;,\;\; 
{\tilde e}_{ab,L} \equiv \frac{e_{ab}^+ - i \, e_{ab}^\times}{\sqrt{2}} \equiv {\tilde e}_{ab,-1} \;. 
\end{equation} 
As shown in Appendix A of \cite{Bartolo:2018qqn}, under a rotation ${\hat n} \to R {\hat n}$, the helicity operators transform as 
\begin{equation}
{\tilde e}_{ab,\lambda} \left( R {\hat n} \right) = {\rm e}^{-2 i \lambda \, \gamma \left[ {\hat n},\, R \right]} \, 
R_{ac} R_{bd} \, {\tilde e}_{cd,\lambda} \left( {\hat n} \right) \;, 
\end{equation} 
where $\gamma$ is a real quantity whose precise expression is not relevant for the present discussion (see 
 \cite{Bartolo:2018qqn} for the precise expression). By combining the last two expressions one finds that 
\begin{equation} 
\sum_{s=+,\times} e_{ab}^s \left( R {\hat n} \right)  e_{cd}^s \left( R {\hat n} \right) = 
R_{aa'} R_{bb'} R_{cc'} R_{dd'} \, \sum_{s=+,\times} e_{ab}^s \left( {\hat n} \right)  e_{cd}^s \left( {\hat n} \right) \;, 
\label{pol-ide2}
\end{equation} 
namely the quantity $\sum_{s=+,\times} e_{ab}^s \left( R {\hat n} \right)  e_{cd}^s \left( R {\hat n} \right) $ 
is a tensor under rotations, while the individual polarization operator is not.

\section{Impact of a nonvanishing phase $\Phi$}
\label{app:Phi}

In this appendix we estimate the impact of neglecting the phase $\Phi$ in eq. (\ref{sti-stj}). Speecifically we show that, once reintroduced in \eqref{Cm}, this phase changes the result only to second order. To see this, we Taylor expand to first order eq. \eqref{Cm} with the phase inserted into it 
\begin{eqnarray} 
\left\langle C_m \right\rangle & = &  \frac{\tau}{T} \int_0^T d t \; {\rm e}^{-i m \omega_e t }
\int_{-\infty}^{+\infty} d f  \, H \left( \vert f \vert  \right) 
\int d^2 {\hat n}{\rm e}^{2 \pi i f {\hat n} \cdot \left( \vec{x}_1(t) - \vec{x}_2(t) \right)}
\sum_{s=+,\times} e_{ab}^s \left( {\hat n} \right) e_{cd}^s \left( {\hat n} \right) \nonumber\\ 
&& \quad\quad\quad\quad  \quad\quad\quad\quad 
\times \sum_{\ell m'} p_{\ell m'} \, Y_{\ell m'} \left( {\hat n} \right)  \, \sum_{O=A,E}  \sum_{O'=A,E}  d^{ab}_O \left( t \right) \, d^{cd}_{O'} \left( t \right) Q_{OO'} \left( f \right)\; \nonumber\\
&\simeq& \frac{\tau}{T} \int_0^T d t \; {\rm e}^{-i m \omega_e t }
\int_{-\infty}^{+\infty} d f  \, H \left( \vert f \vert  \right) 
\int d^2 {\hat n}\left(1+2\pi i f\hat{n}\cdot\Delta\vec x(t)\right)
\sum_{s=+,\times} e_{ab}^s \left( {\hat n} \right) e_{cd}^s \left( {\hat n} \right) \nonumber\\ 
&& \quad\quad\quad\quad  \quad\quad\quad\quad 
\times \,  \sum_{\ell m'} p_{\ell m'} \, Y_{\ell m'} \left( {\hat n} \right)  \, \sum_{O=A,E}  \sum_{O'=A,E}  d^{ab}_O \left( t \right) \, d^{cd}_{O'} \left( t \right) Q_{OO'} \left( f \right) \nonumber\\ 
& \equiv & \left\langle C_m^{(0)} \right\rangle+\left\langle C_m^{(1)} \right\rangle \;, 
\end{eqnarray}
where $C_m^{(i)}$ denotes the order $\Phi^i$ term in the Taylor expansion, and where we have defined $\Delta \vec{x}(t)=\vec{x}_1 (t)-\vec{x}_2(t)$. In the following, we  denote by $L$ the length of this vector. 

We denote the  first order correction of the coefficients \eqref{ga-lm-ab-cd} analogously, 
\begin{eqnarray} 
\gamma_{\ell m,ab,cd}^{(1)}(t) & \equiv & \frac{5}{8 \pi} \int d^2 {\hat n}\,2\pi i f\hat{n}\cdot\Delta\vec x(t) \, Y_{\ell m} \left( {\hat n} \right) 
\sum_P e_{ab}^P \left(  {\hat n} \right) e_{cd}^P \left(  {\hat n} \right) \nonumber\\ 
&& \Rightarrow \;\;\;  \gamma_{\ell m,OO'}^{(1)}(t)  \equiv \gamma_{\ell m,ab,cd}^{(1)}(t) \times d_O^{ab}  \, d_{O'}^{cd} \;. 
\label{ga-lm-ab-cd-1}
\end{eqnarray} 

Differently from the zeroth-order expressions evaluated in the man text, the quantities $\gamma_{\ell m,ab,cd}^{(1)} \left( t \right)$ depend on time. For this reason, we first define a new variable $\alpha\equiv\omega_e\,t$ and we compute $\gamma_{\ell m,ab,cd}^{(1)}(t)$ in a new reference system with coordinates ($\theta'$,$\phi'$), co-rotating with the Earth, and defined in such a way that the vector $\Delta\vec x'$ is oriented along the $z'-$axis \cite{Allen:1996gp}
\begin{equation}
\hat n'\cdot\Delta\vec x'(\alpha)= L \, \cos\theta' \;. 
\end{equation}
We choose the fixed frame such that $\Delta \vec{x}$ has no $y-$component at the initial time. The change of coordinate $\left( \theta,\, \phi \right) \to \left( \theta',\, \phi' \right)$ is then obtained from two consecutive rotations. The first rotation is by an angle $\alpha$ about the $z-$axis, so that, after this rotation, $\Delta \vec{x}$ has vanishing $y-$component at all times. The second rotation is along the new $y-$axis (the one emerging from the first rotation), so to eliminate the $x-$component of $\Delta \vec{x}$. We denote by $\beta$ the angle of this second rotation, and by $R$ the matrix that encodes the product of these two rotations. Accounting for the change of the polarization operator given by eq. (\ref{pol-ide2}) and for the change of the spherical harmonics  under a rotation, we see that the change of variable results in 
\begin{equation}
\begin{cases}
Y_{\ell m}(\theta,\phi)=\sum_{k=-\ell}^lD^{\ell*}_{mk}(\alpha,\beta,0)Y_{\ell k}(\theta',\phi')=\sum_{k=-\ell}^\ell e^{im\alpha}d^\ell_{mk}(\beta)Y_{\ell k}(\theta',\phi') \;, \\
\hat n\cdot\Delta\vec x(\alpha) = L \, \cos\theta' \;,\\
\sum_{s=+,\times} e_{ab}^s \left( {\hat n} \right) e_{cd}^s \left( {\hat n} \right)= 
R_{aa'} \, R_{bb'} \, R_{cc'} \, R_{dd'} \, \sum_{s=+,\times} e_{a'b'}^s \left( {\hat n'} \right) e_{c'd'}^s \left( {\hat n'} \right) \;,\\
d^2\hat n=d^2\hat n' \;, 
\end{cases}
\end{equation}
where the explicit form for the coefficients $d^l_{mk}(\beta)$, which is however irrelevant for our purposes, can be found in \cite{Allen:1996gp}. With these substitution, and setting $u\equiv\cos\theta'$, one finds
\begin{align}\label{eq2}
\gamma_{\ell m,ab,cd}^{(1)}(\alpha)=\frac{5}{8\pi} R_{aa'} \, R_{bb'} \, R_{cc'} \, R_{dd'} \, 
\sum_{k=-\ell }^\ell e^{im\alpha}d^\ell_{mk}(\beta)N^k_\ell\,\int_{-1}^{1}du\,P^k_\ell (u)\,L \, u \, I^k_{a'b'c'd'} \left( u \right) \,, 
\end{align}
where we have used $Y_{\ell k}(\theta',\phi')=\sqrt{\frac{2 \ell +1}{4\pi}\frac{(\ell -k)!}{(\ell +k)!}} P^k_\ell (\cos\theta')e^{ik\phi'}\equiv N_\ell^k P^k_\ell (\cos\theta')e^{ik\phi'}$, and where we have introduced 
\begin{equation}
I^k_{a'b'c'd'} \left( u \right) \equiv\int_0^{2\pi}d\phi' e^{ik\phi'} \sum_{s=+,\times} e_{a'b'}^s \left( {\hat n'} \right) e_{c'd'}^s \left( {\hat n'} \right) \;. 
\end{equation}
One can verify that $I^k_{a'b'c'd'} \left( u \right)$ has parity $\left( - 1 \right)^k$ as $u \to -u$. The associated Legendre polynomials $P^k_\ell \left( u \right)$ have instead parity $\left( - 1 \right)^{k+\ell}$. Therefore, the integrand of the $\int_{-1}^{+1} d u$ integration in (\ref{eq2}) has parity $\left( - 1 \right)^{\ell +1}$. As the integration is on an even domain, it vanishes for even $\ell$, and, as a consequence,  
\begin{equation}
\gamma_{lm,ab,cd}^{(1)}(\alpha)=0  \;\;\; {\rm for \; even \; } \ell \;. 
\end{equation}
This concludes the proof that the terms $\gamma_{0m,ab,cd}^{(0)} ,\, \gamma_{2m,ab,cd}^{(0)} ,\, \gamma_{4m,ab,cd}^{(0)}$ evaluated in the main text, do not receive corrections to linear order in $\Phi$. 
Therefore, our results for these coefficients are accurate up to ${\rm O} \left( \Phi^2 \right)$ corrections. To conclude this appendix, it is worth noting that the same steps outlined here allow to see that only even (odd) powers of $\Phi$ contribute to even (odd) $\ell$ correlators.

\section{Evaluation of the coefficients $\gamma_{\ell m, ab, cd}$}
\label{app:coeff}

In this Appendix we evaluate the coefficients $\gamma_{\ell m, ab, cd}$. We start from eq. (\ref{ga-lm-ab-cd}), where we insert the expression for the spherical harmonics, 
\begin{equation}
Y_{\ell m} \left( \theta, \phi \right)=\sqrt{\frac{2l+1}{4\pi}\frac{(l-m)!}{(l+m)!}} \, 
P^m_\ell \left( \cos\theta \right) \,e^{im\phi} \equiv N_\ell^m \, P^m_\ell \left( \cos\theta \right) e^{im\phi} \;, 
\end{equation}
($\theta$ and $\phi$ are the polar angles that specify the direction ${\hat n}$, and  $P^m_\ell$ are the associated Legendre polynomials) and where we use the second of (\ref{pol-ide1}), to  write 
\begin{align}
\gamma_{\ell m,ab,cd} = \frac{5}{8\pi} & \int_0^{\pi} d \theta \sin \, \theta \, 
N_\ell^m P^m_\ell \left(\cos\theta \right) \int_0^{2\pi} d \phi \, e^{im\phi} \nonumber\\
& \times \Bigg\{ \delta_{ac} \delta_{bd} + \delta_{ad}  \delta_{bc} - \delta_{ab} \delta_{cd}  \nonumber\\
& \quad\quad 
+ 
\delta_{ab} \, {\hat n}_c \, {\hat n}_d + 
\delta_{cd} \, {\hat n}_a \, {\hat n}_b 
-  \left[
\delta_{ac} \, {\hat n}_b \, {\hat n}_d +
\delta_{bd} \, {\hat n}_a \, {\hat n}_c  +
\delta_{ad} \, {\hat n}_b \, {\hat n}_c +
\delta_{bc} \, {\hat n}_a \, {\hat n}_d
\right]  \nonumber\\
& \quad\quad  
+ {\hat n}_a \, {\hat n}_b \, {\hat n}_c \, {\hat n}_d 
\Bigg\} \,. 
\label{gamlmTens}
\end{align}

Let us first discuss the integration over the angle $\phi$. We notice the presence of three structures, characterized by, respectively, zero, two, and four elements ${\hat n}$. 

The terms with no ${\hat n}$ give 
\begin{equation}
\int_0^{2\pi}d\phi \, e^{im\phi} \, \delta_{ab} \delta_{cd} =
\begin{cases} 
2\pi\delta_{ab}\delta_{cd} & \text{if $m=0$} \;, \\
0 & \text{if $|m|>0$} \;,  
\end{cases} 
\label{phi-0}
\end{equation}
(and identically for the other two structures in the second line of eq.  (\ref{gamlmTens})). An explicit evaluation of the terms with two  ${\hat n}$ results in 
\begin{align}
\int_0^{2\pi}d\phi'e^{im\phi'}\delta_{ab} {\hat n_c} {\hat n_d} =\pi\delta_{ab}
\begin{cases}
{\tilde A}_{cd}(\theta) & \text{if $m=0$} \;, \\
{\tilde B}_{cd\pm}(\theta) & \text{if $m=\pm 1$} \;, \\
{\tilde C}_{cd\pm}(\theta) & \text{if $m=\pm 2$} \;, \\
0 & \text{if $|m|>2$} \;, 
\end{cases}
\label{phi-2}
\end{align}
where we have introduced the matrices 
\begin{align}
&{\tilde A}_{cd}(\theta) \equiv \begin{pmatrix}
\sin^2\theta & 0 & 0 \\
0 & \sin^2\theta & 0 \\
0 & 0 & 2\cos^2\theta \\
\end{pmatrix} \;, \\
&{\tilde B}_{cd\pm}(\theta) \equiv \begin{pmatrix}
0 & 0 & \sin\theta\cos\theta \\
0 & 0 & \pm i\sin\theta\cos\theta \\
\sin\theta\cos\theta & \pm i\sin\theta\cos\theta & 0 \\
\end{pmatrix} \;, \\
&{\tilde C}_{cd\pm}(\theta) \equiv \frac 1 2\begin{pmatrix}
\sin^2\theta & \pm i\sin^2\theta & 0 \\ 
\pm i\sin^2\theta & -\sin^2\theta & 0 \\ 
0 & 0 & 0 \\ 
\end{pmatrix} \;. 
\end{align}
 
Finally, an explicit evaluation of the terms with four ${\hat n}$ results in 
\begin{align}
\int_0^{2\pi}d\phi e^{im\phi} 
{\hat n}_a {\hat n}_b 
{\hat n}_c {\hat n}_d = \pi 
\begin{cases}
{\tilde D}_{abcd} + \cos^2 \theta {\tilde E}_{abcd} +  \cos^4 \theta {\tilde F}_{abcd} \;\;  & \text{if $m=0$} \;, \\
\sin \theta \, \cos \theta \left[ {\tilde G}_{abcd\pm} + \cos^2 \theta \,  {\tilde H}_{abcd\pm} \right] \;\;  & \text{if $m=\pm 1$} \;, \\
\left( 1 - \cos^4 \theta \right) {\tilde I}_{abcd\pm} + \cos^2 \theta \, \sin^2 \theta {\tilde J}_{abcd\pm}   \;\;  & \text{if $m=\pm2$} \;, \\
\cos \theta \, \sin^3 \theta \, {\tilde K}_{abcd\pm}  \;\;  & \text{if $m=\pm3$} \;, \\
\sin^4 \theta \, {\tilde L}_{abcd\pm}  \;\;  & \text{if $m=\pm4$} \;, \\
0 & \text{if $|m|>4$} \;,  
\end{cases}
\label{phi-4}
\end{align}
where the matrices $D ,\, \dots \, M_\pm$ are constant (and where the $+$ and $-$ matrices are conjugate of each other). Their explicit form is not illuminating, and we do not report it here. We use  the results (\ref{phi-0}), (\ref{phi-2}), and (\ref{phi-4}) in eq. (\ref{gamlmTens}) and we perform the remaining integration. Considering only the $\theta$ dependences, we have the following integrals (where $x = \cos \theta$) 
\begin{eqnarray}
N_\ell^0 \, \int_{-1}^{1} d x \, P_\ell \left( x \right) &=& \frac{1}{\sqrt{\pi}} \, \delta_{\ell 0} 
\;, \nonumber\\
N_\ell^0 \, \int_{-1}^{1} d x \, P_\ell \left( x \right) \, x^2&=& \frac{1}{3 \sqrt{\pi}} \,\delta_{\ell 0} + \frac{2}{3 \sqrt{5 \pi}} \,\delta_{\ell 2} 
\;, \nonumber\\
N_\ell^0 \, \int_{-1}^{1} d x \, P_\ell \left( x \right) \, x^4&=& \frac{1}{5 \sqrt{\pi}}  \,\delta_{\ell 0} +  \frac{4}{7 \sqrt{5 \pi}}  \,\delta_{\ell 2}  +  \frac{8}{105 \sqrt{\pi}}  \,\delta_{\ell 4} 
\;, 
\end{eqnarray}
for the coefficient $m=0$, 
\begin{eqnarray}
N_\ell^{\pm1} \, \int_{-1}^{1} d x \, P_\ell^{\pm 1} \left( x \right) x \sqrt{1-x^2} &=& \mp \sqrt{\frac{2}{15 \, \pi}}  
\, \delta_{\ell 2} 
\;, \nonumber\\
N_\ell^{\pm1} \, \int_{-1}^{1} d x \, P_\ell^{\pm 1} \left( x \right) x^3 \sqrt{1-x^2} &=& 
\mp \frac{1}{7} \sqrt{\frac{6}{5 \pi} }  \, \delta_{\ell 2} \mp  \frac{4}{21 \, \sqrt{5 \, \pi}} \, \delta_{\ell 4} 
\;, 
\end{eqnarray}
for the coefficient $m=\pm1$, 
\begin{eqnarray}
N_\ell^{\pm2} \, \int_{-1}^{1} d x \, P_\ell^{\pm 2} \left( x \right) \left( 1 - x^2 \right) &=& 
2 \sqrt{\frac{2}{15 \, \pi}} \, \delta_{\ell 2}
\;, \nonumber\\
N_\ell^{\pm2} \, \int_{-1}^{1} d x \, P_\ell^{\pm 2} \left( x \right) \left( 1 - x^4 \right) &=& 
\frac{16}{7} \sqrt{\frac{2}{15 \, \pi}}  \, \delta_{\ell 2} + \frac{4}{21} \sqrt{\frac{2}{5 \, \pi}}  \, \delta_{\ell 4} 
\;, \nonumber\\
N_\ell^{\pm2} \, \int_{-1}^{1} d x \, P_\ell^{\pm 2} \left( x \right) x^2 \left( 1 - x^2 \right) &=& 
\frac{2}{7} \sqrt{\frac{2}{15 \, \pi}}  \, \delta_{\ell 2} + \frac{4}{21} \sqrt{\frac{2}{5 \, \pi}}  \, \delta_{\ell 4} 
\;, 
\end{eqnarray} 
for the coefficient $m=\pm2$, 
\begin{eqnarray}
N_\ell^{\pm3} \, \int_{-1}^{1} d x \, P_\ell^{\pm3} \left( x \right) x \left( 1-x^2 \right)^{3/2} &=& 
\mp \frac{4}{3 \sqrt{35 \, \pi}}  \, \delta_{\ell 4} \;, 
\end{eqnarray} 
for the coefficient $m=\pm2$, and, finally 
\begin{eqnarray}
N_\ell^{\pm4} \, \int_{-1}^{1} d x \, P_\ell^{\pm4} \left( x \right)  \left( 1-x^2 \right)^2 &=& 
\frac{8}{3} \, \sqrt{\frac{2}{35 \, \pi}}  \, \delta_{\ell 4} \;,  
\end{eqnarray} 
for the coefficient $m=\pm3$. 

Inserting these results, together with eqs. (\ref{phi-0}), (\ref{phi-2}), and (\ref{phi-4}), in eq. (\ref{gamlmTens}), we obtain the expressions given in eq. (\ref{gamma-result}) of the main text.

\end{document}